\documentclass[reprint,superscriptaddress,amsmath,amssymb,aps,longbibliography]{revtex4-2}
\usepackage{graphicx}
\usepackage{overpic} 

\usepackage{graphicx}
\usepackage{dcolumn}
\usepackage{bm}
\usepackage[caption=false]{subfig} 
\usepackage{mwe} 
\usepackage{amsmath}
\usepackage{amssymb}
\usepackage{ulem} 
\usepackage[dvipsnames]{xcolor}
\usepackage{soul}   

\begin{document}

\preprint{APS/123-QED}

\title{Shape matters: Body dynamics underlies efficient jumping}

\author{Marc Su\~n\'e}
\email{marc.sune@mansfield.ox.ac.uk}
\affiliation{%
 Mathematical Institute, University of Oxford, Oxford OX2 6GG, United Kingdom
}
\affiliation{
 Mansfield College, University of Oxford, Oxford OX1 3TF, United Kingdom
}%

\author{Lucas Selva}
\affiliation{%
 Mathematical Institute, University of Oxford, Oxford OX2 6GG, United Kingdom
}%

\author{Crist\'obal Arratia}
\affiliation{
 Nordita, Royal Institute of Technology and Stockholm University, Hannes Alfvéns väg 12, SE-106 91 Stockholm, Sweden
}%

\author{John Wettlaufer}
 \email{john.wettlaufer@yale.edu}
 \affiliation{
 Nordita, Royal Institute of Technology and Stockholm University, Hannes Alfvéns väg 12, SE-106 91 Stockholm, Sweden
}%
\affiliation{
 Yale University, New Haven, Connecticut 06520, USA\\
 and Nordita, Royal Institute of Technology and Stockholm University, Hannes Alfvéns väg 12, SE-106 91 Stockholm, Sweden
}%

\author{Dominic Vella}%
 \email{dominic.vella@maths.ox.ac.uk}
\affiliation{%
 Mathematical Institute, University of Oxford, Oxford OX2 6GG, United Kingdom
}%

\begin{abstract}
Many small animals, particularly insects, use power-amplification to generate rapid motions, such as jumping, that would otherwise  be impossible given the standard power density of muscle. A common framework for understanding this power amplification is  Latch-Mediated, Spring Actuated (LaMSA) jumping, in which a spring is slowly compressed, latched in its compressed state and the latch released to allow jumping. Motivated by the jumps of certain insect larvae, we consider an external latching mechanism via adhesion to a substrate that is quickly released for jumping.
We show that the rate at which this adhesion is lost is crucial in determining the efficiency of jumping and, indeed, whether jumping occurs at all. As well as showing how release rate should be chosen to facilitate optimal jumping, our analysis underscores the importance of the interaction between latch-release dynamics and the elastic deformation of the jumper for power amplification, thereby providing new insight into the post-latch control of jumping.
\end{abstract}

\date{\today}

\maketitle

Power amplification is an important strategy for generating fast motion in both the engineering and natural sciences: an archer stores energy slowly in the string of a bow allowing the arrow to fly faster when released than would have been the case if it had simply been thrown \cite{GalantisWoledge03}. Similarly, plants such as the Venus flytrap \cite{Forterre2005} and insects such as the  `click' beetle family \cite{BolminSocha21} are able to generate rapid motions by storing energy slowly in an effective spring that is released quickly when required.

A key ingredient for actuating power-amplified motions in elastic structures is the decoupling of the low-power input from the high-power output. A `latch' -- the archer’s fingers holding the arrow and bowstring in a drawn bow -- is an exemplar decoupling mechanism that enables the temporal separation of low-power energy storage and high-power actuation.
Latch mechanisms are usually part of the elastic body itself as for the~click beetles~\cite{BolminSocha21} and many invertebrates~\cite{Patek2015}, but also in artificial jumping systems~\cite{Yang2012,WangWang23}.
Their geometric/structural properties and the dynamics of their removal (e.g. latch removal velocity and unlatching duration) have proven to be pivotal for controlling the transformation of elastic potential energy into kinetic energy in `Latch-Mediated, Spring Actuated' (or LaMSA) mechanical systems~\cite{KagayaPatek16,IltonBhamla18,Divi2020,Divi2023,HyunOlberding23}. However, the latched object such as a jumper is usually modelled as a simple spring-mass system.

\begin{figure*}[htbp]
\centering
\begin{overpic}[width=0.36\linewidth]{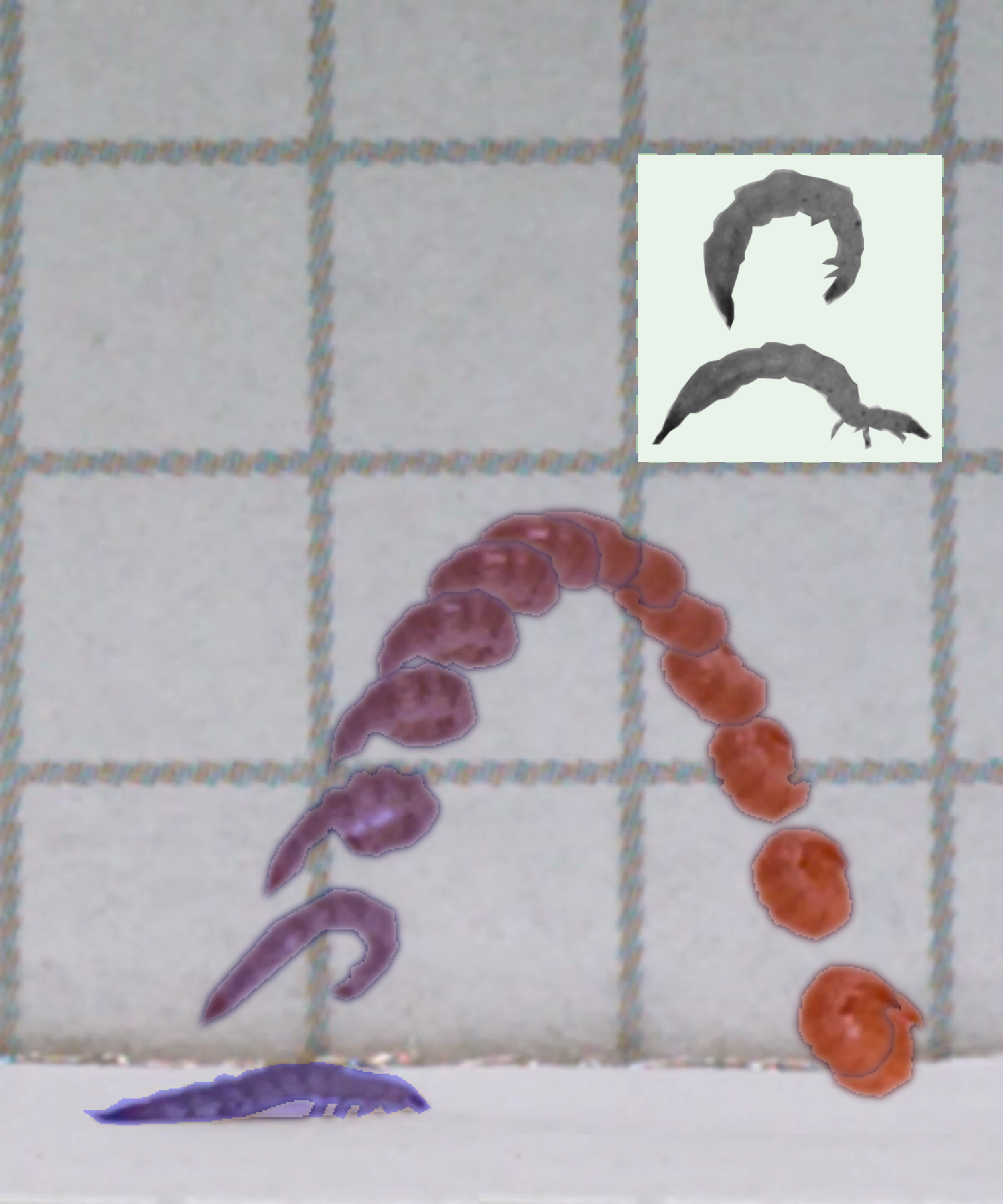}\put(3,92){\textbf{A}}\end{overpic}\hfill
\begin{overpic}[width=0.288\linewidth]{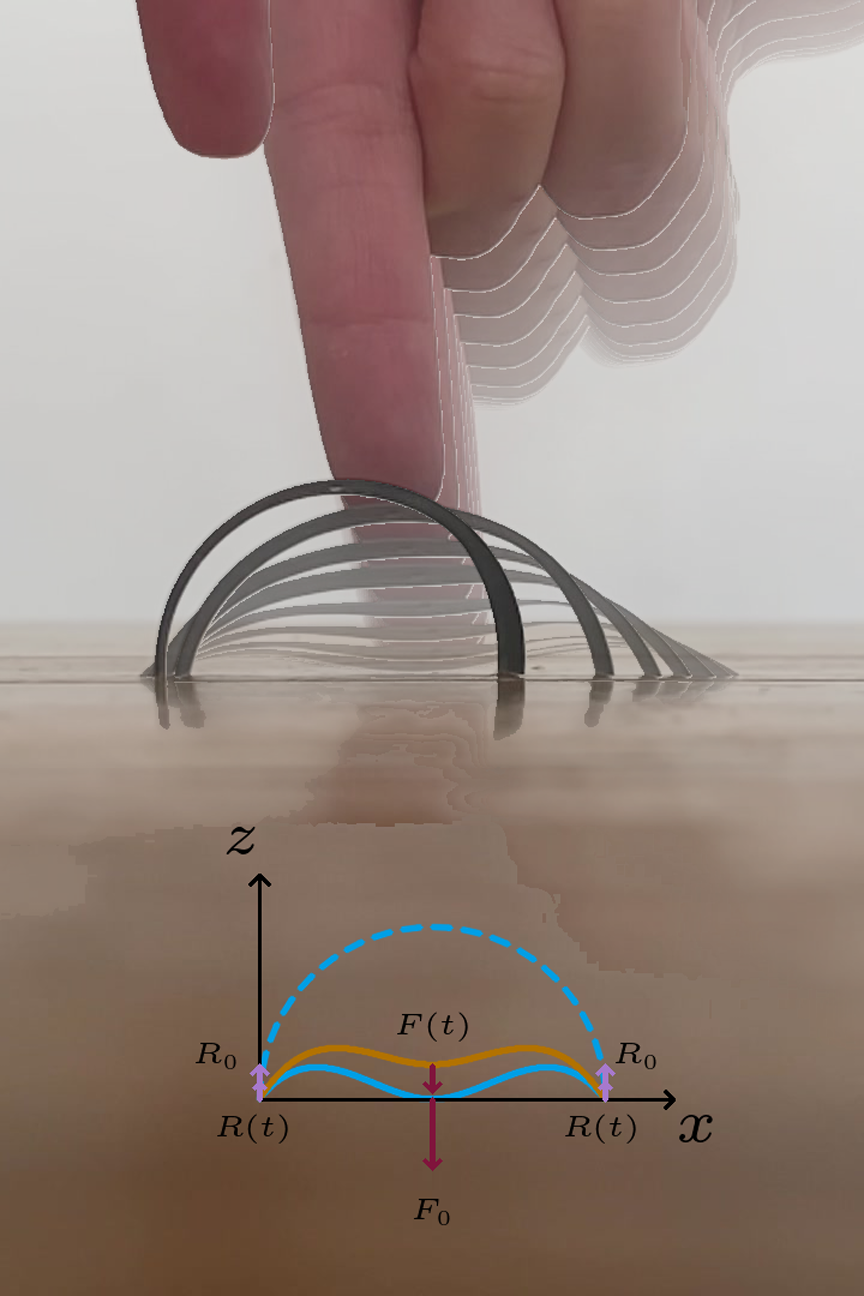}\put(3,92){\textbf{B}}\end{overpic}\hfill
\begin{overpic}[width=0.27\linewidth]{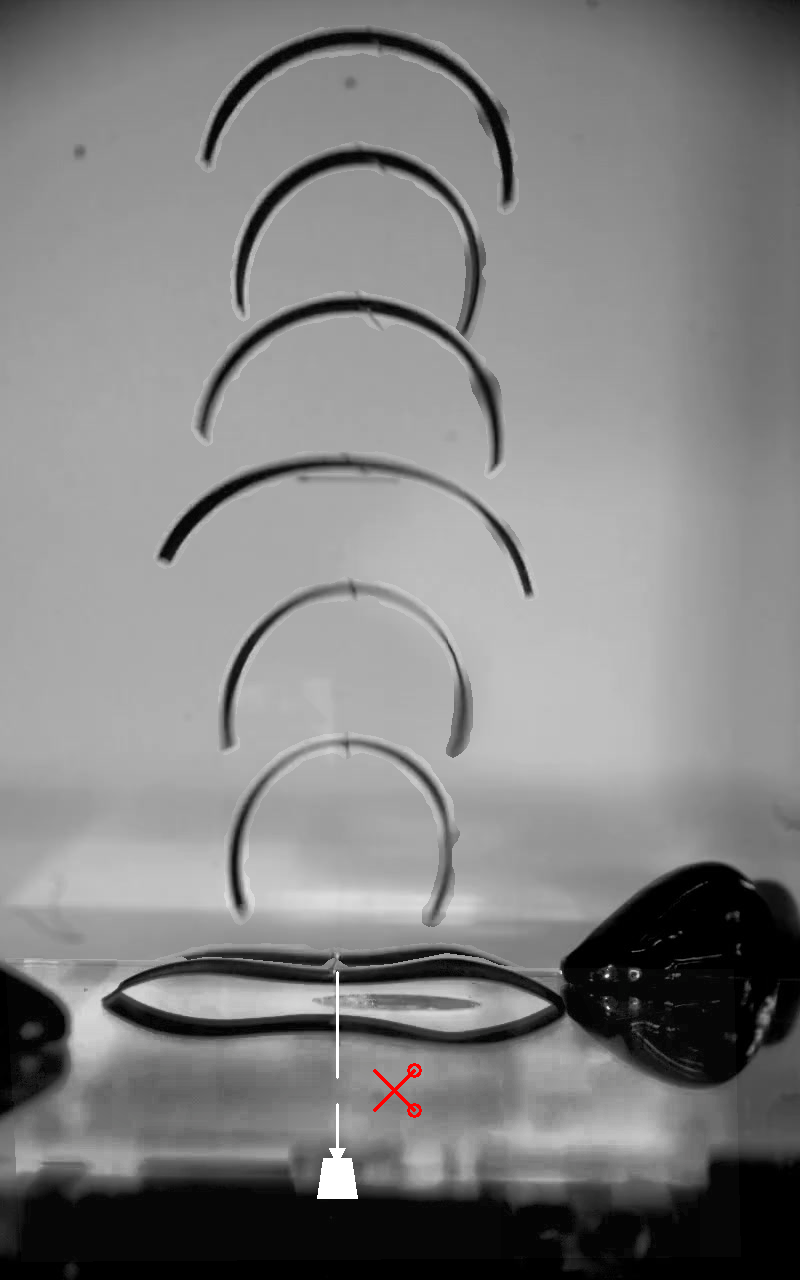}\put(3,92){\textbf{C}}\end{overpic}

\caption{\textbf{(A)} Jumping behavior in larval beetles (\emph{Laemophloeus biguttatus}) --- equally spaced frames show instants up to $100~\text{ms}$ after the initiation of jumping. Inset: flattened shape when the latch is mounted (below), and the curved state after the latch has been released and the larva has jumped (above). (Adapted from~\cite{BertoneGibson22}.)
\textbf{(B-C)} A curved elastic strip is initially indented at its center by a pressing finger \textbf{(B)} and a hanging weight (white pictogram) \textbf{(C)}. A slow release does not result in jumping (\textbf{(B)}, equally timed frames  up to $867~\text{ms}$ after first motion), while quick release does (\textbf{(C)} with equally timed frames up to $64~\text{ms}$ after first motion).
  Unlatching times in the time composite images \textbf{(A-C)} are $t_u=(5.5, 683, 1)~\text{ms}$.
  Inset in panel \textbf{(B)}: Schematic representation of the shell in the $xz$-plane in three different configurations: in the absence of external forces, other than its own weight (dashed blue),  loaded at the center (solid blue), and partially released loading (orange).}
\label{fig:1}
\end{figure*}

In other cases another body (most often a substrate) serves as an anchor and the effective latch is the connection between the jumper and substrate. This unusual `external latch' behavior is exhibited by some larval beetles~\cite{BertoneGibson22}. It is understood that the larva attaches its legs to the substrate and then arches its body dorsally by muscle contraction. Because of the attachment to the substrate, the arching of the larva's back does not significantly change the shape but effectively stores elastic energy in the elastic components of its body (the spring) while its legs grip the substrate (thereby forming the latch). When the legs begin to lose contact with the substrate, the constraint imposed by the latch is released and the body is able to  return to its (now curved) natural state. In the process, the stored elastic energy is transferred to kinetic energy.
 The combination of relaxation to a curved shape and jumping is shown in Fig.~\ref{fig:1}A.
 
The same function of decoupling power \emph{in} and \emph{out} can also be observed in snapping instabilities: a spherical cap of finite thickness may be everted at low power but quickly snaps to its natural state upon release~\cite{TaffetaniJiang18}.
Applications that harness snap-through to generate power-amplified motions include jumping toys~\cite{PandeyMoulton14} as well as plants~\cite{Forterre2013} and animals~\cite{SmithYanega11,KuanChiu20}. 

In addition to the decoupling mechanism, jumping requires a rigid surface to generate the reaction force for takeoff. This poses a complex problem that combines elastic deformation with a dynamically evolving contact region.
In this context, the power-amplified jumping of snapping spherical caps has been examined with soft robots and finite-element simulations~\cite{GorissenMelancon20,AbeHashiguchi25,AbeSano26}, and scaling analyses based on quasi-static assumptions and energy conservation~\cite{AbeHashiguchi25,AbeSano26}. However, a detailed study of how useful work is generated, and the role of elastic deformation at the body scale, is still lacking.

Motivated by  the example of snapping shells and inspired by the external force-controlled latch mechanism in larval beetles~\cite{BertoneGibson22}, we investigate the efficiency of power-amplified jumping with a geometrical model of a naturally curved elastic strip. We consider confining the strip by poking it down at its center and then releasing it (mimicking the legs' grip on the substrate and its release). Figure \ref{fig:1}B shows a superposition of the states that the arch goes through when the confining force is removed relatively slowly --- the strip goes from being approximately flat to being naturally curved, but does not appear to lose contact with the substrate. However, if the confining force is removed more quickly --- for example by hanging a weight below the arch on a thread that can readily be cut (Fig.~\ref{fig:1}C) --- then the strip jumps from the substrate. Clearly, the strip stores elastic potential energy when it is indented that may be converted to kinetic energy. More surprisingly perhaps, the efficiency of this conversion depends on how quickly the indenting force is released, raising the question of how release speed changes the efficiency of this energy conversion.

The jumping larvae studied by~\cite{BertoneGibson22} exhibit a large scatter in the efficiency of this mechanism --- the coefficient of variation of the kinetic energy at take off is about $25\%$ for jumps by the same individual. Given the role of release rate in the biomimetic example of a curved strip, it is natural to examine its role in externally latched jumping as a means of controlling the jump characteristics even once the latch has been set.

We consider a two-dimensional model in which a naturally curved, linearly elastic and inextensible cylindrical shell lies on the $xz$-plane with both ends at $z=0$ (see the dashed blue curve in the  schematic --- lower inset --- of Fig.~\ref{fig:1}B). The shell is in contact with a flat, rigid substrate perpendicular to the gravitational field (acceleration $g$). We assume that contact only occurs at these two ends, simplifying the analysis considerably. The shell has natural curvature  $\kappa_0$, volumetric mass density $\rho$, Young's modulus $E$, thickness $h$ and length $\ell$, with $h\ll \ell$.

Initially, an indentation force  $f(t=0)=f_0$ (per unit width) is applied at the center of the shell, confining it to lie almost parallel to the substrate (solid blue curve in the schematic of Fig.~\ref{fig:1}B). This force serves as the motor (increasing and storing elastic energy within the deformed shell for $t<0$). However, once the indenting force reaches $z=0$, $f$ plays the role of the latch, keeping the deformed shell confined --- the shell is in equilibrium due to the reaction force of the substrate on the edges. For time $t>0$ the indentation force is gradually released (orange curve in the schematic of  Fig.~\ref{fig:1}B) with characteristic unlatching time $t_u$.

In the following we non-dimensionalize lateral lengths by $\ell$, heights by $\ell^2|\kappa_0|$, mass (per unit width) by $\rho h\ell$, times by $(\rho h \ell^4/EI)^{1/2}$, and vertical forces (per unit width) by $EI|\kappa_0|/\ell$. (Here,  $I=h^3/12$ is the shell's second moment of area\footnote{See Supplemental Material for more details of the calculations described in the main text.}) We use upper case letters to denote dimensionless versions of the corresponding dimensional quantity, e.g.~$X=x/\ell$, $Z=z/(|\kappa_0|\ell^2)$. In the limit of small, planar deflections, the small, time-dependent, transverse displacements $Z(X,T)$ of the shell satisfy the dynamic beam equation~\cite{HowellKozyreff2008}:
\begin{equation}
    \frac{\partial^4Z}{\partial X^4}+\frac{\partial^2Z}{\partial T^2}+ H \frac{\partial Z}{\partial T}+1/\beta=0.
    \label{eq:beam}
\end{equation} 
Here we neglect the tension within the shell, and hence the role of friction with the substrate; this is justified within the small angle approximation where the work done against friction is negligible because of the small horizontal distance moved by the contact points \cite{Yoshida2020}. This approximation further simplifies the analytic treatment of the problem because, while the shell remains in contact with the substrate, the boundary conditions are fixed at $X=0,1$: $Z(0,T)=Z(1,T)=0$ and $\left.\partial^2Z/\partial X^2 \right|_{X=0,1}=\kappa_0/|\kappa_0|$.

In Eq.~\eqref{eq:beam}, two dimensionless groups are introduced:
\begin{align}
    \beta=\frac{EI|\kappa_0|}{\rho g h \ell^2} \quad\textrm{and}\quad H= \frac{2\eta \ell^2}{(EI\rho h)^{1/2}}.
    \label{eq:groups}
\end{align} These measure the relative importance of bending forces to the gravitational force and the strength of damping relative to bending forces, respectively. $\eta$ is the translational drag coefficient of the shell that we fit from the oscillations in the experiments of Fig.~\ref{fig:1}C (the decay of oscillations in our experiments suggests that $H\approx 2$ \cite{Note1}). The typical value of $\beta$ in the experiments is approximately 10, while we estimate $\beta\approx1$ for the larvae of Fig.~\ref{fig:1}A \cite{Note1};  we must therefore account for gravity in what follows.

The applied force $F(T)$ enters via a point loading condition at $X=1/2$ --- we assume that there is a discontinuity in the shear force there (or, equivalently, a Dirac $\delta$-function external pressure). We assume a simple model for a force-controlled latch with characteristic unlatching time $t_u$, and therefore have that
\begin{equation}
    \left[\frac{\partial^3Z}{\partial X^3}\right]_{X=1/2^-}^{X=1/2^+}=    F_0\exp(-T/\tau),
    \label{eqn:ForceBC}
\end{equation} where $\tau=t_u (E I/(\rho h \ell^4))^{1/2}$ and $F_0=f_0 \ell/(E I |\kappa_0|)$.

We take as the initial condition that the indentation force is precisely that required to make the center touch the substrate, i.e.~$Z(X,T=0)=Z_0(X)$ where $Z_0(X)$ is the solution to the linearized equilibrium problem
\begin{equation}
    \frac{{\rm d}^4 Z_0}{{\rm d} X^4}+\frac{1}{\beta}=0,
    \label{eq:statics}
\end{equation}
with analogous boundary conditions at $X=0,1$ and $Z_0(X=1/2)=0$. (The combination of these conditions yields a unique solution to \eqref{eq:statics} \cite{Note1}.) The value of $F_0$ in \eqref{eqn:ForceBC} is determined from the resulting discontinuity in the shear force. Finally, the shell is assumed to start from rest, so that~$\partial Z/\partial T(X,T=0)=0$.

The dynamic beam equation \eqref{eq:beam} with the  boundary, initial and continuity conditions discussed above can be solved analytically, though the result is involved \cite{Note1}.

Both the jumping of the larvae and our experiments reveal a broad range of jumping behavior from failed attempts to leave the ground to effective (and impressive) jumps; there is clearly large variability in the efficiency of this mechanism. We rationalize this behavior in terms of the three dimensionless parameters in the model: $\beta,\tau$ and $H$.
 We begin by noting that the scaled reaction force of the ground on the edge of the shell, $R$, can only be positive --- if this reaction force becomes negative then the shell has lost contact with the ground, and the shell has jumped; $R$ may be calculated from the solution $Z(X,T)$ as
\begin{equation}
    R(T)=\left.\frac{\partial^3Z}{\partial X^3}\right|_{X=0}.
\end{equation}

\begin{figure}[htbp]
  \centering
  \begin{overpic}[width=0.9\linewidth]{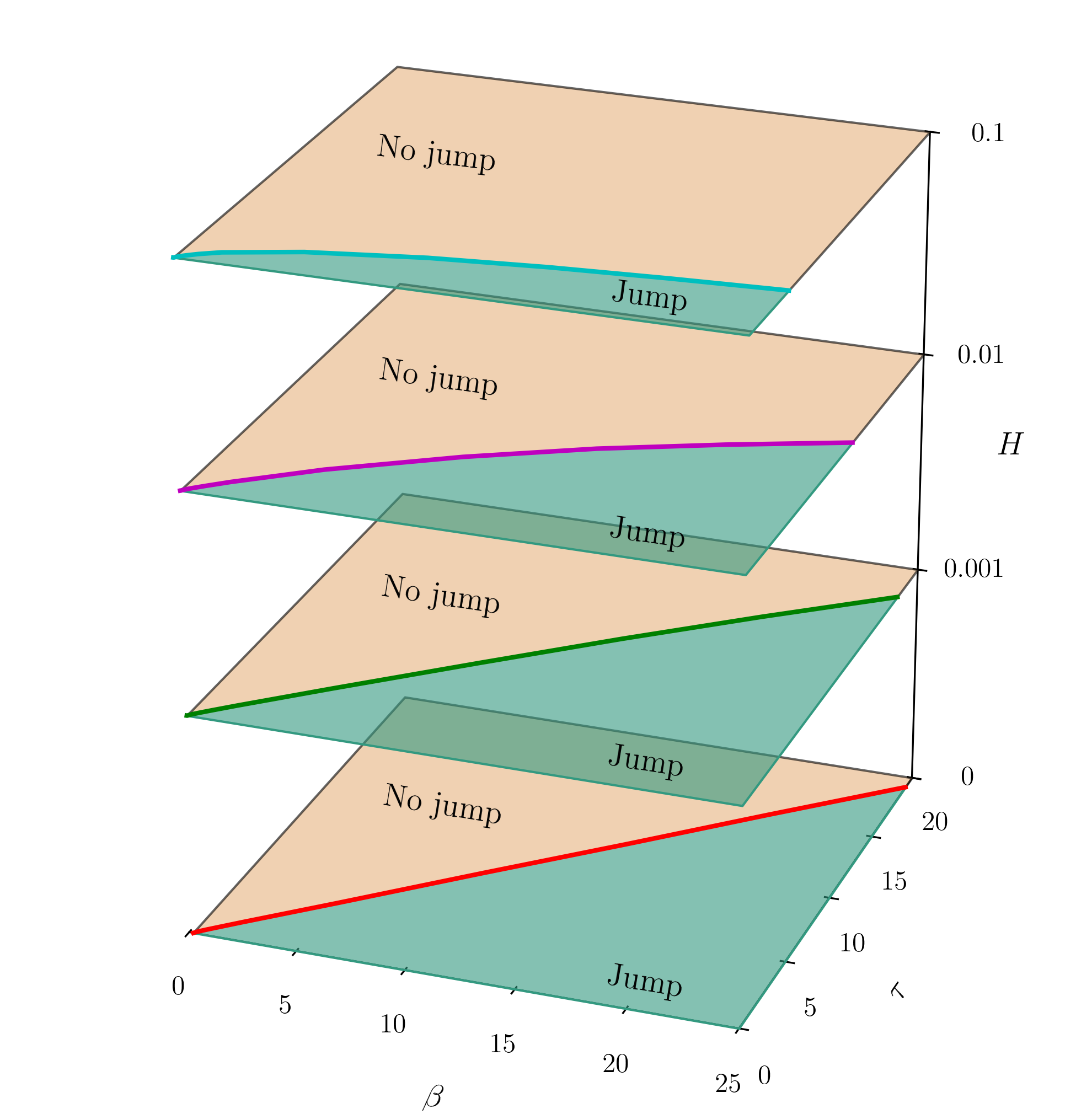}\put(3,92){\textbf{A}}\end{overpic}
  
  \vspace{4em}
  
  \begin{overpic}[width=0.9\linewidth]{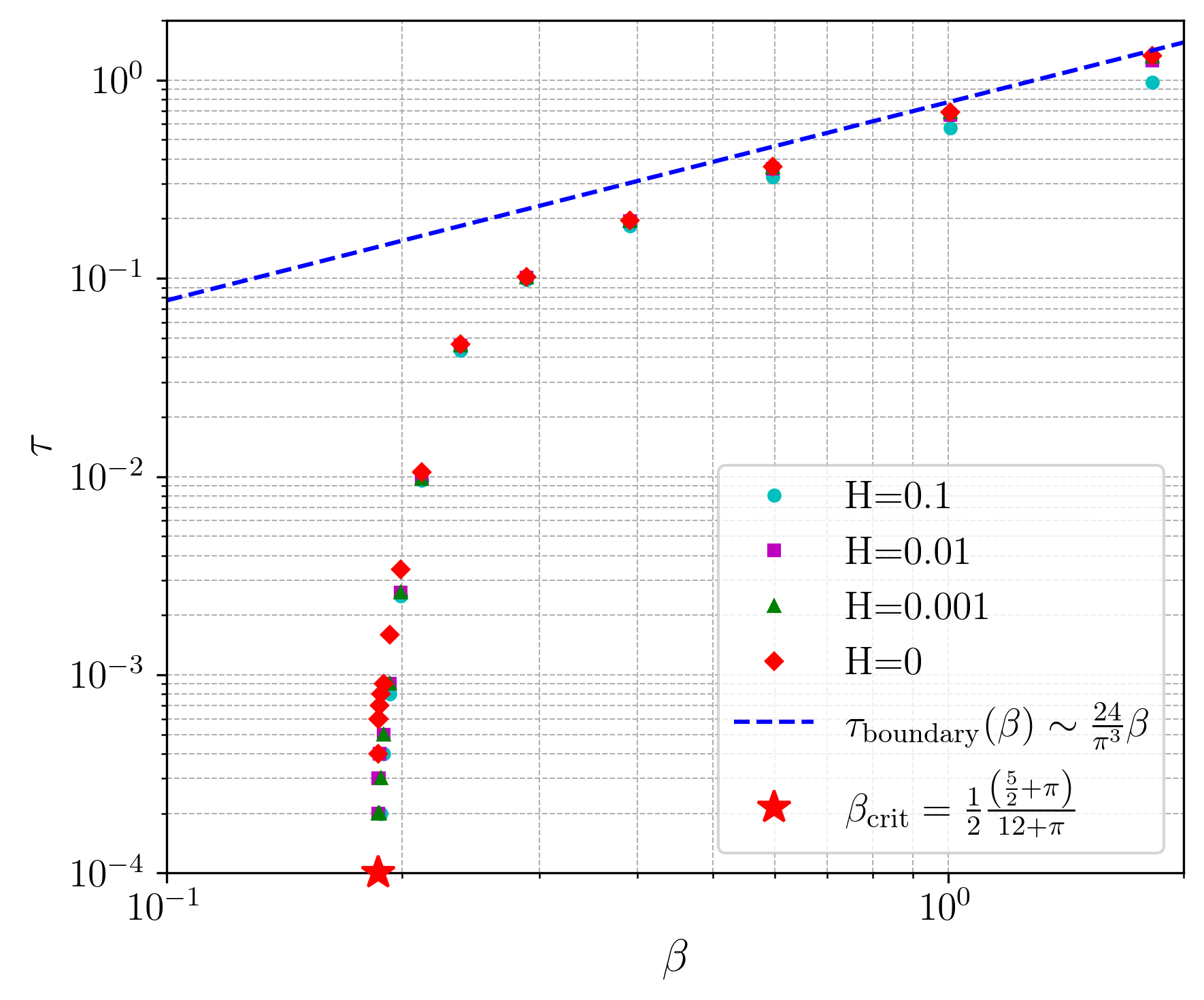}\put(3,85){\textbf{B}}\end{overpic}
\caption{\label{fig:2} (\textbf{A}) Phase diagram of the dimensionless elasticity ($\beta$) and latch release times ($\tau$) for which jumping is possible with different values of the damping $H$.
(\textbf{B}) Phase boundaries at different values of parameter $H$ and small $(\beta,\tau)$. The color scheme of the boundaries in (\textbf{A}) is the same as the points in (\textbf{B}).}
\end{figure}

We then build the phase space of the failure/success in jumping from the analytical solution of the linearized model, finding the smallest positive root of  $R(T)=0$ numerically using the \texttt{Chebfun} package~\cite{Chebfun} in \texttt{Matlab}.  The results in Fig.~\ref{fig:2}A show that the shell jumps for larger $\beta$ (light, stiff shell), smaller $\tau$ (quick release) and $H$ (small dissipation), but also that it fails to jump for smaller $\beta$ (heavy shell), larger $\tau$ (slow release) and $H$ (large dissipation).

In undamped conditions ($H=0$), for light, stiff shells (i.e.~away from the small $\beta$ limit), the phase boundary in the $(\beta,\tau)$ plane is  linear. An asymptotic approximation of the analytical solution  --- in the limit of large $\tau$ and large $T/\tau$  --- allows us to calculate an approximation of $R(T)$ 
and successively an estimate of the phase boundary $\tau_{\text{boundary}}(\beta)\sim \frac{24}{\pi^3}\beta$  \cite{Note1} that shows good agreement with numerical results for $\beta>1$ (see Fig.~\ref{fig:2}B). This boundary becomes noticeably sublinear  as the dissipation $H$ increases: jumping requires a quicker unlatching for  more strongly damped systems. Moreover, regardless of the damping $H$, a strong nonlinear deviation of the phase boundary also occurs for small $\beta$  (Fig.~\ref{fig:2}B). The phase boundary crosses the abscissa at $\beta\sim\beta_{\text{crit}}$, where $\beta_{\text{crit}}=\frac{1}{2}\frac{\left(\frac{5}{2}+\pi\right)}{12+\pi}\approx 0.186$ is an estimation (by asymptotic arguments \cite{Note1}) of the minimal $\beta$ for jumping. Intuitively, the existence of a non-zero minimal $\beta$ reflects the fact that a sufficiently heavy shell will bend due to its own weight and hence naturally satisfy the condition $Z_0(X=1/2)=0$ at zero external force and hence cannot jump. (A first estimate of $\beta_{\text{crit}}$ by this argument gives $\beta_{\text{crit}}=5/48\approx 0.104$ \cite{Note1}.)

The phase space in Fig.~\ref{fig:2} quantifies the intuition that a faster unlatching mechanism facilitates jumping. We therefore expect that a jumping mechanism should be designed to ensure the fastest possible unlatching. However, the larvae studied by \cite{BertoneGibson22} seem to contradict this basic principle as they grip the substrate with all their legs, presumably resulting in a slower latch release. To understand this apparent paradox we examine the maximum dimensional latch force $f_0$ --- i.e.~the attaching force between a larva's legs and the substrate --- and the stored energy (bending + gravitational).
This requires an estimate of the effective Young's modulus of the larva, that we obtain by using the time scale of the return to the naturally curved state once the jump has been initiated, and comparing this to theoretical work on the curling of a ribbon by \cite{Callan-JonesBrun12} (we estimate $E_{\rm eff}\sim 10\mathrm{~kPa}$, and a natural curvature $|\kappa_0^{\max}|=1.69\mathrm{~mm}^{-1}$ \cite{Note1}.

\begin{figure}[t]
  \centering
    \includegraphics[width=\linewidth]{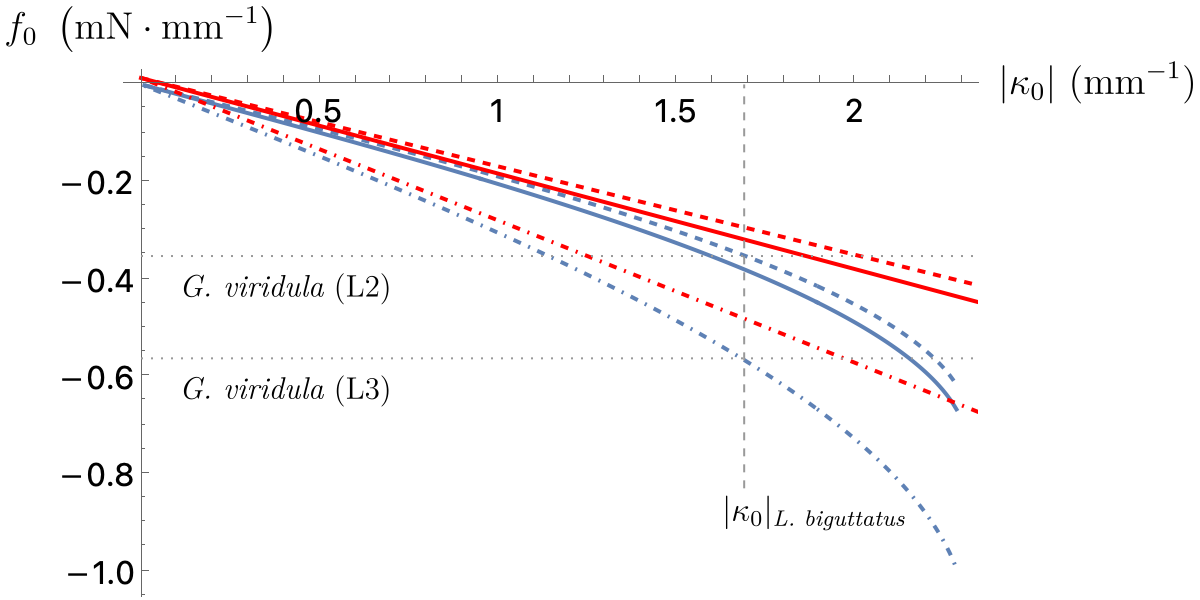}
  \caption{\label{fig:Force}Latch force of a fully indented shell as a function of the shell's natural curvature given by the linearized theory (in red) and the non-linear model (in blue). Re-dimensionalization was done using geometrical parameters from {\em Laemophloeus biguttatus}~\cite{BertoneGibson22}  and  $E\sim 10~\mathrm{kPa}$ (solid curves; estimated using theory of~\cite{Callan-JonesBrun12}); $E\sim 9.25~\mathrm{kPa}$ (dashed curves; estimated from attachment force of 2nd instar of \emph{G. viridula}~\cite{ZurekGorb15});  $E\sim 14.9~\mathrm{kPa}$ (dot--dashed curves; estimated from attachment force of 3rd instar of \emph{G. viridula}~\cite{ZurekGorb15}). (Here `instar' refers to different developmental stages of insects, between molting of the exoskeleton.)
  }
  \label{fig:3}
\end{figure}

The results of $f_0$ --- both by the discontinuity at $X=1/2$ in the third derivative of the linear solution $Z_0(X)$ of \eqref{eq:statics} and a geometrically nonlinear model \cite{Note1} --- show (Fig.~\ref{fig:3}) that the magnitude of the latch force increases approximately linearly as the curvature of the shell in its natural state increases --- even when geometrical nonlinearities are accounted for. However, the stored energy shows a superlinear increase with \emph{natural} curvature \cite{Note1}: the bending energy of the shell increases with deviations of the (instantaneous) curvature from the shell’s natural curvature. (The bending energy is proportional to the square of the difference between these two curvatures \cite{Note1}.) Given that the shell is always close to flat in its latched configuration, shells with larger natural curvature will have a bigger deviation of curvature (and hence higher bending energy) than shells with a smaller natural curvature.  See, for example, the inset in Fig.~\ref{fig:1}A, which shows the flattened shape of the larva when the latch is set (below), and the curved state after the latch has been released (above).

Assuming that each pair of legs is able to produce a similar force, and noting that they are located nearby one another (and hence can be modelled by a single point force), using three pairs of legs to anchor on the surface --- as the larvae in \cite{BertoneGibson22} --- should increase the maximum natural curvature obtained before release by a factor of 3 and the energy stored by a factor of 9.

Using the linear results in Fig.~\ref{fig:3} and measuring the maximum natural curvature $\kappa_0^{\max}$ from the videos in~\cite{BertoneGibson22}, we estimate the maximum latch force that can be generated by \emph{Laemophloeus biguttatus} by approximating its body by a rod of radius $\sim 0.6\mathrm{~mm}$. We find a maximum force of around $0.2\mathrm{~mN}$ --- this is consistent with experimental results on other insect larvae \cite{ZurekGorb15}, which show that the legs are typically able to supply a maximum adhesive force of $0.3-0.6\mathrm{~mN}$.
Finally, we hypothesize that once the maximum force is exceeded at one point and that once adhesion between leg and substrate fails, then the remaining legs are successively detached.

A key parameter of importance in the space of LaMSA mechanisms is the efficiency of the release mechanism: what percentage of the elastic energy that is initially stored is realized as translational kinetic energy upon release? Ref.~\cite{Yang2012} reported this for jumping elastic hoops and found that it is a universal quantity, around $57\%$, independent of the material properties of their system. While it is difficult to ascribe a precise value of $\tau$ to  their experimental technique, it is very small, and agrees well with the predicted efficiency when $\tau\approx0$.  Here, we readily calculate the efficiency of jumping, defined as
\begin{equation}
    \mathcal{E}=\frac{K_{\rm trans}}{E_0}\times 100\%
\end{equation} directly from the analytical solution.
\begin{figure}[!t]
\centering
\includegraphics[width=\columnwidth]{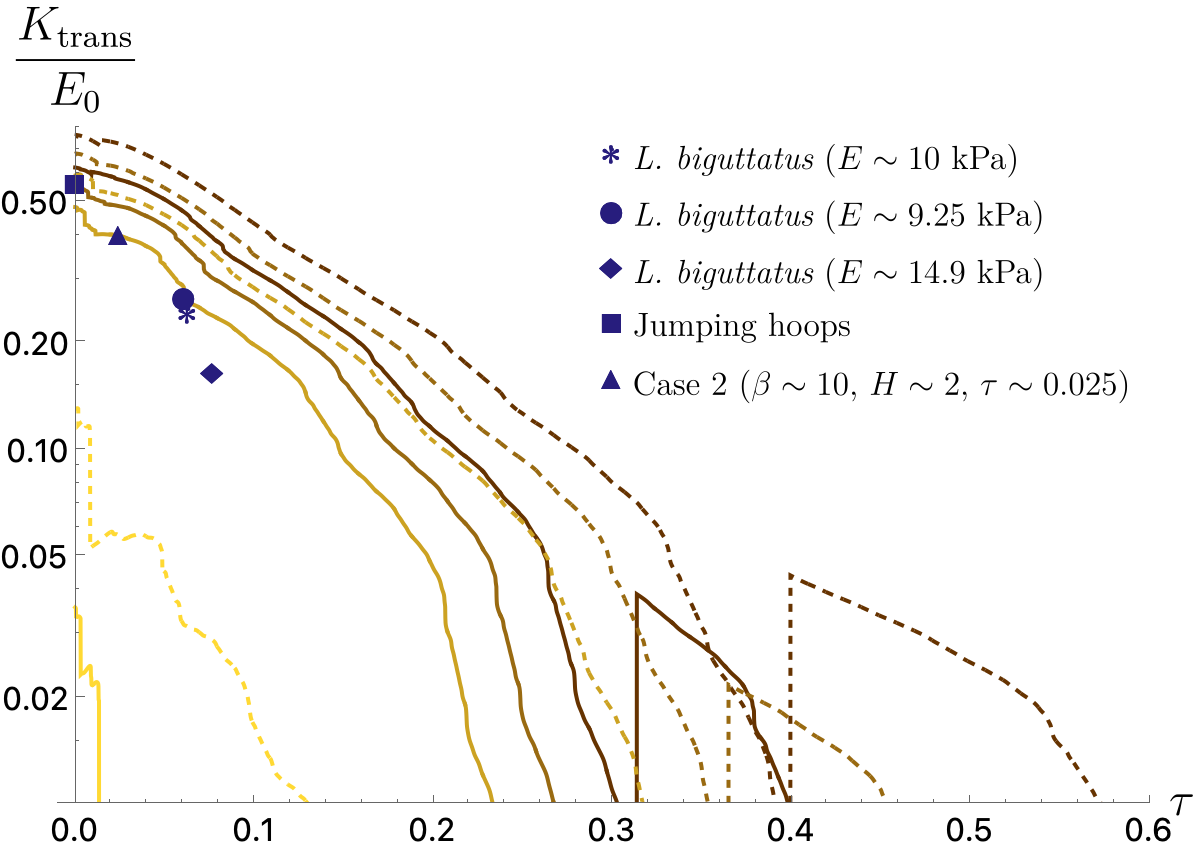}
\caption{Energy efficiency of jumping of a curved shell (the ratio of translational kinetic energy at $T=T_{\text{off}}$ to stored energy at $T=0$) as a function of the latch lifting time scale. Theoretical predictions from linearized theory are depicted by solid ($\beta=1$) and dashed ($\beta=10$) curves. Scaled values of the efficiency for \emph{L. biguttatus}~\cite{BertoneGibson22}, jumping hoops~\cite{Yang2012}, and our experiments with a curved strip are  included as symbols.  Curve colors range from dark brown (small $H$) to yellow (large $H$) for $H=0.1,1,2,10$.}
\label{fig:efficiency}
\end{figure}
Our results for $\mathcal{E}(\tau)$ (for a variety of values of $\beta$ and $H$) in Fig.~\ref{fig:efficiency} suggest an exponential decrease of the efficiency of jumping with $\tau$, as well as the existence of certain \emph{pessimal} values of $\tau$ where the efficiency drops even more dramatically. We attribute this to a destructive interference between the imposed decay of $F(t)$ and body dynamics (i.e. the natural relaxation time of the shell), which is consistent with the re-entrant efficiency observed for lighter, stiffer shells.  (Using this idea, we estimate these pessimal values  using asymptotic methods \cite{Note1}.)

Crucially, the exponential decay of the efficiency is not a mere consequence of the exponential force release \eqref{eqn:ForceBC}. Indeed, for a simplified 1D mass-spring model (following Ref.~\cite{AguilarLesov12}), different release protocols give  decay of jumping efficiency markedly slower than exponential \cite{Note1}.
This result underscores the importance of integrating force control of latch release with a geometric model of the jumper’s elastodynamic behavior.
The observed sensitive dependence of the efficiency on the release time scale $\tau$ suggests that the dynamic deformation of the strip itself accentuates this -- a feature that is not captured in the LaMSA formalism~\cite{Divi2020,Divi2023}.

We have developed a geometrical theory that integrates latch-mediated energy release and useful work generation to amplify power, thereby providing a geometrical description of jumping that unifies the most relevant aspects of LaMSA systems~\cite{Divi2020,Divi2023,HyunOlberding23} and jumping robots~\cite{Yang2012,GorissenMelancon20,WangWang23,AbeHashiguchi25,AbeSano26}.
Our theory describes the observed efficiency of the jumps of insect larvae, as well as providing a mechanism for the wide distribution of observed jump efficiencies: an exponential decay means that small changes in the time scale of latch release, $\tau$, can lead to significantly less efficient jumps --- in agreement with the wide variability in the efficiency of jumps that were reported by \cite{BertoneGibson22}. This may pose both a  challenge and an opportunity to insect larvae: while the timing of release is important, it also gives a new means of modifying the jump \emph{after} the energy has been stored and the spring `primed'. This may also help to focus the design criteria for controlling jumps in robots \cite{WangWang23}, for example, providing another setting of insect-inspired robotics \cite{Liang2025}.\\

\noindent
This work was partially supported by the UK Engineering and Physical Sciences Research Council via Grant No.~EP/W016249/1 and Grant No.~EP/Y027949/1 (M.S.~and D.V.), and by the Swedish Research Council under Grant No.~638-2013-9243 (M.S.~and J.S.W.). For the purpose of Open Access, the authors will apply a CC BY public copyright license to any Author Accepted Manuscript version arising from this submission. M.S.~thanks Maria Bruna for helpful comments.

\bibliography{jumping}

\end{document}


\title{\textbf{Supplementary Information for ``Shape matters: Body dynamics underlies efficient jumping''}}

\author{Marc Su\~n\'e}
\author{Lucas Selva}
\author{Crist\'obal Arratia}
\author{John Wettlaufer}
\author{Dominic Vella}%

\maketitle

In this Supplementary Information we present further details of the two-dimensional model of the cylindrical shell (\S~\ref{sec:model}), its resolution both analytically and numerically (\S~\ref{sec:solutions}), and the analysis of the jumping behavior (\S~\ref{sec:jumping}). We also briefly describe the experiments with a naturally curved strip (\S~\ref{sec:experiments}). Finally, we study the jumping efficiency of a LaMSA mechanism in a simplified 1D mass-spring system (\S~\ref{sec:mass-spring}).

\section{\label{sec:model} 2D model: cylindrical cap and latch mechanism}

\begin{figure*}[htbp]
  \centering
      \includegraphics[width=0.85\linewidth]{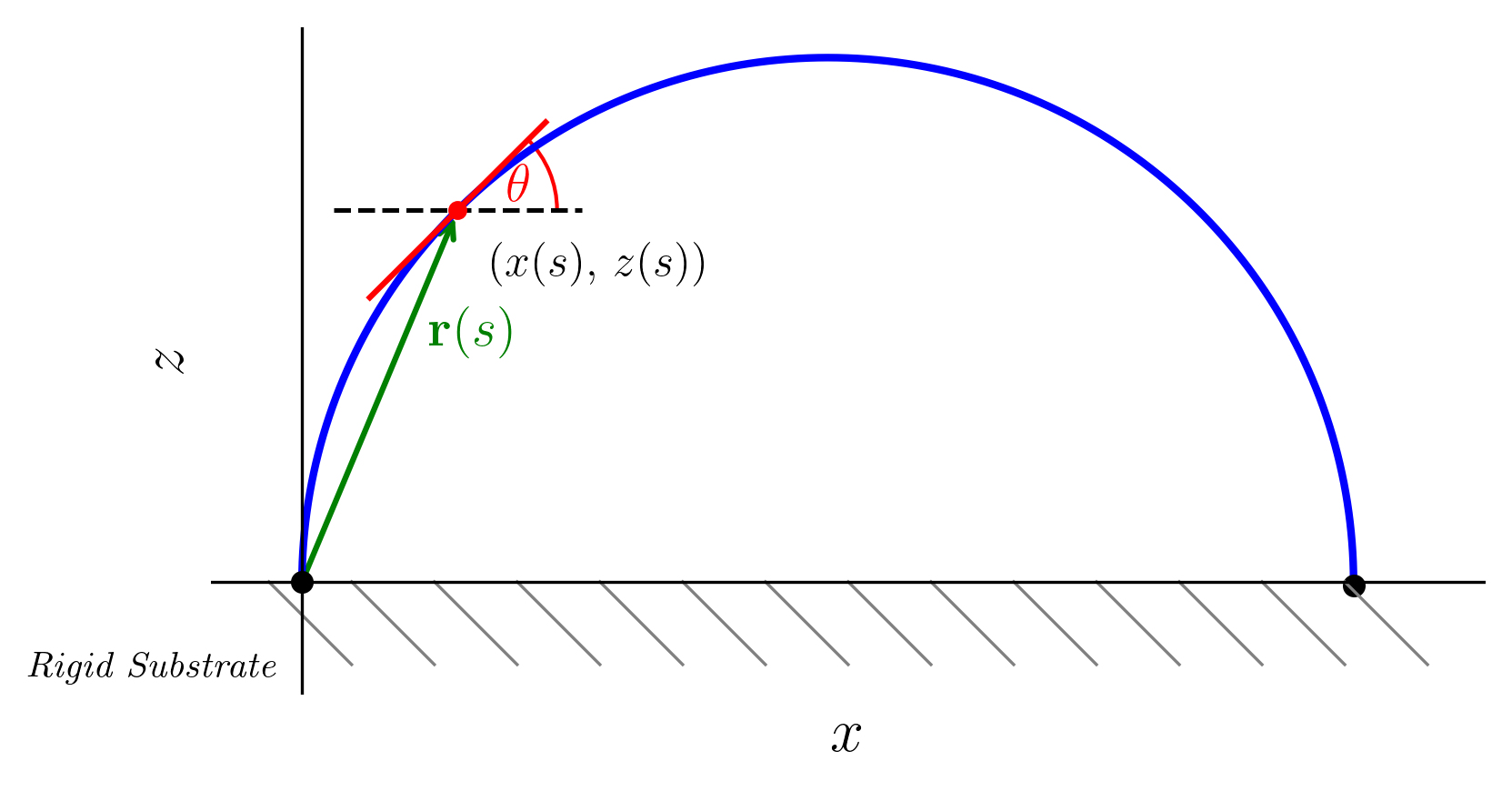}
  \caption{Schematic representation of the cylindrical cap in the $xz$-plane.
  }
  \label{fig:cylindrical_shell}
\end{figure*}

We consider the planar elastic unshearable deformations of a naturally curved, linearly elastic, homogeneous, isotropic and inextensible cylindrical cap. The cylindrical cap forms a shell that lies in the $xz$-plane with both ends at $z=0$, as shown in \Fig{\ref{fig:cylindrical_shell}}. The plane $z=0$ represents a flat, rigid substrate and is perpendicular to the gravitational field (with acceleration $g$). We shall assume that contact only occurs at the two ends of the cylindrical cap.

The shape of the shell is parametrized by the arc length $s$ measured along the centerline from the left end, $s\in[0,\ell]$, with tangent angle $\theta$ --- the angle the tangent vector makes with the horizontal axis. In the absence of any external forces, the shell has constant curvature $\kappa_0$  so that its undeformed shape, $\theta=\theta_0(s)$, satisfies
\begin{equation}
\frac{\mathrm{d}\theta_0(s)}{\mathrm{d}s}= \kappa_0.
    \label{eq:curvature}
\end{equation}

The centerline of the shell, $\mathbf{r}(s)=[x(s),z(s)]$, is given by 
\begin{equation}
    \frac{\partial\mathbf{r}}{\partial s}=\cos{\theta}\,{\mathbf e}_x+\sin{\theta}\,{\mathbf e}_z,
\end{equation}
where ${\mathbf e}_x$ and ${\mathbf e}_z$ are unit vectors respectively in the $x$ and $z$ directions.

The shell has volumetric mass density $\rho$, Young's modulus $E$, second moment of area $I=\frac{h^3}{12}$, thickness $h$, and length $\ell$, with $h\ll \ell$. Because we assume a two-dimensional model of the shell deformations we ignore any dependence along the width (i.e. in the $y$-axis).

\subsection*{\label{sec:beam_equation}Dynamic beam equation}
\subsubsection*{\label{sec:variations}Variational principle}
The equations describing the evolution of the shell's shape can be derived from a variational principle~\cite{AudolyPomeau2010}. We consider the system's generalized coordinates ${\bf q}=(x,z,\theta)$.
We seek to minimize the action  $\mathcal{S}[{\bf q}]=\int_{t_1}^{t_2}\mathrm{d}t\,\mathcal{L}\Big({\mathbf q},\dot {\mathbf q}(t),t\Big)$, with $t$ denoting time, dots denoting differentiation with respect to $t$, and $\mathcal{L}\Big({\mathbf q},\dot {\mathbf q}(t),t\Big)$ denoting the Lagrangian of the system:
\begin{equation}
    \mathcal{L}\Big({\mathbf q},\dot {\mathbf q}(t),t\Big)=\mathcal{T}-\mathcal{V}-\lambda_x(s,t)\left(\frac{\partial x}{\partial s}-\cos\theta\right)-\lambda_z(s,t)\left(\frac{\partial z}{\partial s}-\sin\theta\right)-L\,\lambda_{g}(t)\int_0^L\mathrm{d}s\,\sin\theta.
    \label{eq:lagrangian}
\end{equation}
The system has kinetic energy (per unit width) $\mathcal{T}$, considering translational and rotational contributions,
\begin{equation}
    \mathcal{T}=\frac{1}{2}\int\mathrm{d}z\int\mathrm{d}x \,\rho\left(\dot{x}^2+\dot{z}^2+z^2 \dot{\theta}^2\right)=\frac{1}{2}\rho\int_0^\ell\mathrm{d}s\left[h\left(\dot{x}^2+\dot{z}^2\right)+\frac{h^3}{12}\dot{\theta}^2\right];
    \label{eq:kinetic}
\end{equation}
and potential energy (per unit width) $\mathcal{V}$, which includes gravitational and bending contributions,
\begin{equation}
    \mathcal{V}=\int_{0}^\ell\mathrm{d}s\left[\frac{1}{2}\, E I\left(\frac{\mathrm{d}\theta}{\mathrm{d}s}-\kappa_0\right)^2+\rho g h\,z(s)\right].
\end{equation}
We enforce the inextensibility of the shell (i.e.~the parameter $s$ is the true arc length) and (local) geometric consistency by using  Lagrange multipliers $\lambda_{x,z}(s,t)$. $\lambda_g(t)$ represents the Lagrange multiplier used to enforce the global constraint that both end-points of the shell must have the same  vertical position.

Using the Euler--Lagrange equations with this variational principle we obtain the balance of angular momentum:
\begin{equation}
    -\rho I \frac{\partial^2\theta}{\partial t^2}+E I\frac{\partial^2\theta}{\partial s^2}-\rho g h (\ell-s)\,\cos\theta-\lambda_x(s,t)\,\sin\theta+\lambda_z(s,t)\,\cos\theta-\lambda_g(t)\,\cos\theta=0;
    \label{eq:balance_angular_momentum}
\end{equation}
and the balance of forces in each direction:
\begin{align}
    \rho h\frac{\partial^2 x}{\partial t^2}-\frac{\partial\lambda_x(s,t)}{\partial s}= \,&0,\quad\text{and}\label{eq:balance_force_x}\\
    \rho h\frac{\partial^2 z}{\partial t^2}-\frac{\partial\lambda_z(s,t)}{\partial s}=\,&0.\label{eq:balance_force_z}
\end{align}
Considering freely rotating and freely (horizontally) moving ends --- recall that we assumed fixed vertical positions for the ends --- the boundary conditions are:
\begin{align}
    \frac{\partial\theta(s,t)}{\partial s}\Big|_{s=0,\ell}=\kappa_0,\quad\text{and}
    \quad\lambda_x(s,t)\Big|_{s=0,\ell}=0.\label{eq:bcs}
\end{align}
These boundary conditions implicitly assume the neglect of friction between the shell's endpoints and the ground, we will  discuss this assumption further later.

\subsubsection*{\label{sec:dissipation}Dissipation}
To model the jumping behavior of slender elastic shells we introduce dissipation into \Eqs{\ref{eq:balance_angular_momentum}, \ref{eq:balance_force_x} \& \ref{eq:balance_force_z}}. The shell is confined in the $xz$-plane and hence has three degrees of freedom: rotations about an axis perpendicular to the plane, and translations along the $x,z$ directions. By virtue of the small-slope approximation --- that we will introduce later --- these translations are approximately the axial and horizontal translations of a slender body. Considering the resistance of the surrounding fluid on the moving shell, slender-body theory~\cite{Cox1970} predicts that the horizontal drag coefficient (we omit here its expression in terms of the geometrical parameters of the shell) is approximately twice the axial drag, $\eta$. Thus, viscous damping of translation is modeled by adding damping terms $\eta \,\partial x/\partial t$ and $2\eta \,\partial z/\partial t$ to the balance of forces \Eqs{\ref{eq:balance_force_x} \& \ref{eq:balance_force_z}}, respectively. The viscosity of the surrounding fluid also induces a viscous torque (per unit width) $\tau\propto \gamma\,\partial\theta/\partial t$ in the balance of angular momentum~\Eq{\ref{eq:balance_angular_momentum}}, whose rotational damping coefficient, $\gamma\propto \ell\, h\,\eta$, is also given by slender-body theory, see e.g.~\cite{Batchelor1970}.

\subsubsection*{\label{sec:scaling}Scaling and slender-body approximation ($h/\ell\ll 1$)}
We nondimensionalize all lengths by $\ell$, heights by $\ell^2|\kappa_0|$, mass (per unit width) by $\rho h\ell$, times by $(\rho h \ell^4/EI)^{1/2}$, horizontal forces (per unit width) by $EI/\ell^2$ and vertical forces (per unit width) by $EI|\kappa_0|/\ell$. 
To denote dimensionless versions of the variables in our model we use upper case versions of the symbols used to denote the corresponding dimensional quantity, e.g.~$S=s/\ell$, $T=t(EI/\rho h\ell^4)^{1/2}$, but $Z=z/(|\kappa_0|\ell^2)$ etc. Assuming the slender-body approximation ($h/\ell\ll 1$), the scaled equations for the balance of angular momentum \Eq{\ref{eq:balance_angular_momentum}} and the balances of forces \Eqs{\ref{eq:balance_force_x} \& \ref{eq:balance_force_z}}, including dissipative terms, are:
\begin{align}
    &\frac{\partial^2\theta}{\partial S^2}-\frac{\gamma}{(EI\rho h)^{1/2}}\frac{\partial\theta}{\partial T}-\frac{\rho g h \ell^3}{EI}(1-S)\cos{\theta}-\Lambda_{X}\sin{\theta}+\ell|\kappa_0|\Lambda_{Z}\cos{\theta}-\ell|\kappa_0|\Lambda_g\cos{\theta}=0;\label{eq:balance_angular_momentum_sc}\\
    &\frac{\partial^2 X}{\partial T^2}+\eta\frac{\ell^2}{(EI\rho h)^{1/2}}\frac{\partial X}{\partial T}-\frac{\partial\Lambda_{X}}{\partial S}=0,\quad\text{and}\label{eq:balance_force_x_sc}\\
    &\frac{\partial^2 Z}{\partial T^2}+2\eta\frac{\ell^2}{(EI\rho h)^{1/2}}\frac{\partial Z}{\partial T}-\frac{\partial\Lambda_{Z}}{\partial S}=0;\label{eq:balance_force_z_sc}
\end{align}
and the boundary conditions \Eq{\ref{eq:bcs}}:
\begin{align}
    \frac{\partial\theta}{\partial S}\Bigg|_{S=0,1}=\kappa_0 \ell\quad\text{and}
    \quad \Lambda_{X}(S,T)\Big|_{S=0,1}=0.\label{eq:bcs_sc}
\end{align}

\subsubsection*{\label{sec:linear}Small-slope approximation: linearization}
We linearize the balance of angular momentum, \Eq{\ref{eq:balance_angular_momentum_sc}},  by assuming that all angles are small, i.e.~$\theta\approx \ell|\kappa_0|\partial Z/\partial X \ll 1$. We may then approximate $\cos{\theta}\approx 1$, $\sin{\theta}\approx\tan{\theta}= \ell|\kappa_0|\partial Z/\partial X$, $\partial S\approx\partial X$ etc.)
We also assume that $\kappa_0\ell\ll1$ so that the cylindrical cap is shallow and hence that the edges lie at $X=0,1$ --- this is valid near the initial condition when the shell is pushed flat. 
\begin{equation}
    \frac{\partial^3 Z}{\partial X^3}-\frac{\gamma}{(EI\rho h)^{1/2}}\frac{\partial^2 Z}{\partial T\partial X}-\frac{\rho g h \ell^2}{EI|\kappa_0|}(1-X)+\Lambda_{Z}-\Lambda_g=0,
    \label{eq:balance_forces}
\end{equation}
where we have substituted $\Lambda_{X}=0$ (which may be derived from \Eq{\ref{eq:balance_force_x_sc}}, noting that $\partial/\partial S\approx\partial/\partial X$, the inextensibility of the shell, and the boundary conditions \Eq{\ref{eq:bcs_sc}}).
Differentiating \Eq{\ref{eq:balance_forces}} with respect to $X$, and substituting $\partial\Lambda_{Z}/\partial X$ from \Eq{\ref{eq:balance_force_z_sc}}
, we obtain the dynamic beam equation~\cite{HowellKozyreff2008}
\begin{equation}
    \frac{\partial^4 Z}{\partial X^4}-\Gamma\frac{\partial^3 Z}{\partial X^2\partial T}+\frac{1}{\beta}+H\frac{\partial Z}{\partial T}+\frac{\partial^2 Z}{\partial T^2}=0;
    \label{eq:linearized_beam}
\end{equation}
where the three dimensionless groups
\begin{equation}
    \beta = \frac{E I |\kappa_0|}{\rho g h \ell^2},\quad H = \frac{2\eta \ell^2}{(E I \rho h)^{1/2}}\quad \text{and}\quad \Gamma=\frac{\gamma}{(E I \rho h)^{1/2}}
\end{equation}
measure  the relative importance of bending forces to gravitational force, the damping frequency --- the relative importance of viscous forces per unit area to bending forces --- and the ratio of the viscous torque to restoring torque by the shell's bending stiffness, respectively. By virtue of the slender-body approximation, and recalling that $\gamma\propto \ell\, h\,\eta$ (as discussed in ``\emph{Dissipation}''), we have that
\begin{equation}
    \frac{\Gamma}{H}\propto\frac{h}{\ell}\ll 1\nonumber,
\end{equation}
and hence we neglect the rotational damping in the the dynamic beam equation \Eq{\ref{eq:linearized_beam}}, which becomes 
\begin{equation}
    \frac{\partial^4 Z}{\partial X^4}+\frac{1}{\beta}+H\frac{\partial Z}{\partial T}+\frac{\partial^2 Z}{\partial T^2}=0.
    \label{eq:linearized_beam_damped}
\end{equation}
The corresponding linearized boundary conditions to \Eq{\ref{eq:bcs_sc}} and the constraint imposed by $\Lambda_g$ (i.e. $\int_0^1\mathrm{d}S \sin{\theta}=0$)  are
\begin{eqnarray}
    \frac{\partial^2 Z}{\partial X^2}\Bigg|_{X=0,1}=\frac{\kappa_0}{|\kappa_0|}\quad\text{and}\quad Z(X,T)\Big|_{X=0,1}=0.
    \label{eq:bcs_linear}
\end{eqnarray}

\subsection*{\label{sec:latch}Latch and release}
\begin{figure*}[htbp]
  \centering
      \includegraphics[width=0.85\linewidth]{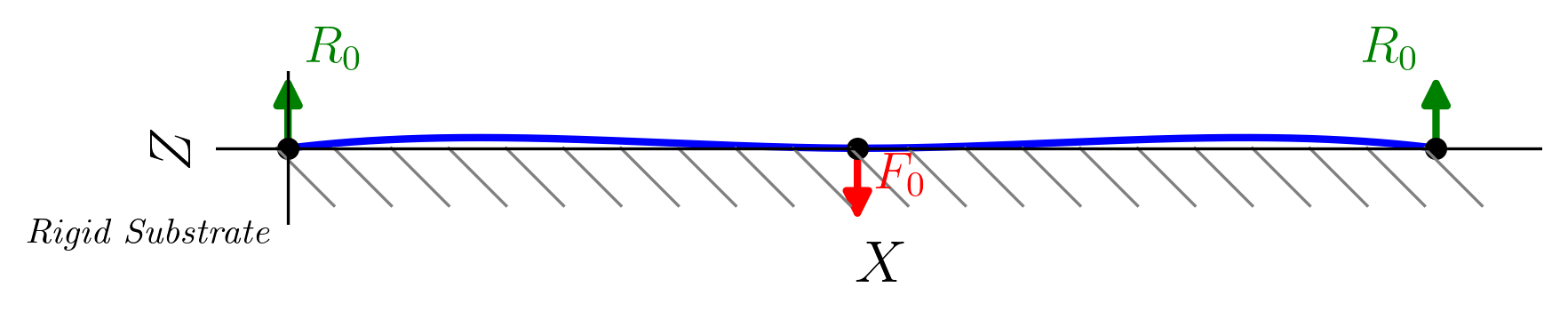}
  \caption{Schematic representation of the deformed shell by an indenting force $F_0$ applied at $X=0$ that serves as a latch.
  }
  \label{fig:latch}
\end{figure*}

We model the latch mechanism and its release as an indenting force $F(T)$ (dimensionless force per unit width, scaled by $EI|\kappa_0|/\ell$) at the center of the shell ($X=0.5$).
Initially, $F(0)\equiv F_{0}$ and this force keeps the deformed shell confined such that its center and endpoints are at the same height, i.e. $Z=0$, see \Fig{\ref{fig:latch}}. The shell is in equilibrium due to the reaction force $R_0$ (dimensionless force per unit width, scaled by $EI|\kappa_0|/\ell$) of the substrate on the edges. The equilibrium in the $Z$ direction (assuming the small-slope approximation in \Eq{\ref{eq:balance_force_z_sc}} and extending its domain so that the endpoints are included in the balance, i.e. $X\in[0,1]$) is then
\begin{equation}
    \delta(X)\,2 R_0+\delta(X-1)\,2 R_0+\delta(X-1/2)\,F_0-\frac{\partial\Lambda_{Z}}{\partial X}=0,
    \label{eq:equilibrium_latch}
\end{equation}
where we have used Dirac delta functions --- represented by $\delta$ --- to introduce the indenting force and the corresponding reaction by the substrate.

We break the domain of \Eq{\ref{eq:equilibrium_latch}} in two subdomains: $X_L\in[0,1/2]$ and $X_R\in[1/2,1]$. (We use $L,R$ notation to label quantities in each domain.) We obtain from the integral of \Eq{\ref{eq:equilibrium_latch}} over $X_L,X_R$:
\begin{eqnarray}
    \Lambda_{Z}^{L}(X=0)=-R_0,\quad \Lambda_{Z}^{L}(X=1/2)=\frac{1}{2}F_0,\nonumber\\ \Lambda_{Z}^{R}(X=1/2)=-\frac{1}{2}F_0,\quad
    \text{and}\quad \Lambda_{Z}^{R}(X=1)=R_0.
    \label{eq:forces}
\end{eqnarray}
Next we use these results and the static version of \Eq{\ref{eq:balance_forces}} to evaluate $\partial^3Z/\partial X^3$ at $X=1/2^-,1/2^+$ to obtain that the latch induces a discontinuity of the third derivative at $X=1/2$:
\begin{equation}
    \left[\frac{\partial^3 Z(X,0)}{\partial X^3}\right]_{X=1/2^-}^{X=1/2^+}=F_0.
    \label{eq:discont_3rd_deriv}
\end{equation}

Therefore, the shape of the shell --- when the latch is applied --- will be given by the solution of the time-independent version of \Eq{\ref{eq:linearized_beam_damped}} with boundary conditions given by \Eq{\ref{eq:bcs_linear}} and the latch constraint $Z_0(X=1/2)=0$, which we have just seen  induces a discontinuity in the third derivative of the shape.

For time $T>0$ the indenting force is gradually released. We assume a simple model for release with dimensionless characteristic unlatching time $\tau$,
\begin{equation}
    F(T)=F_0\exp(-T/\tau).
    \label{eq:latch}
\end{equation}
The corresponding balance of forces eq.~\eqref{eq:balance_force_z_sc} (assuming again the small-slope approximation and the extended domain, $X\in[0,1]$) is:
\begin{equation}
    \delta(X)\,2 R(T)+\delta(X-1)\,2 R(T)+\delta(X-1/2)\,F(T)+\frac{\partial^2 Z}{\partial T^2}+2\eta\frac{\ell^2}{(EI\rho h)^{1/2}}\frac{\partial Z}{\partial T}-\frac{\partial\Lambda_{Z}}{\partial X}=0,
    \label{eq:equilibrium_release}
\end{equation}
and hence the discontinuity in the third derivative is now:
\begin{equation}
    \left[\frac{\partial^3 Z(X,T)}{\partial X^3}\right]_{X=1/2^-}^{X=1/2^+}=F_0\exp{(-T/\tau)}.
    \label{eq:discont_3rd_deriv_dyn}
\end{equation}

The latch release phase is thus modelled by the dynamic beam equation \Eq{\ref{eq:linearized_beam_damped}} equipped with the set of boundary conditions \Eq{\ref{eq:bcs_linear}} and the discontinuity in the third derivative \Eq{\ref{eq:discont_3rd_deriv_dyn}}. Initial conditions are provided by the solutions of the static problem, i.e.~with the latch applied, and assuming that the shell is at rest when the release phase starts, i.e. \begin{equation}
    \left.\frac{\partial Z}{\partial T}\right|_{T=0}=0.
    \label{eqn:ZdotIC}
\end{equation} In the next section we compute these solutions.

\section{\label{sec:solutions}Analytic solutions}

\subsection*{\label{sec:statics}Statics: The shape of a latched shell}

\subsubsection*{Linearized model}
Assuming small deflections, the equilibrium shape of the shell when the latch is  applied (see \Fig{\ref{fig:latch}}) is denoted by $Z_0(X)$ and satisfies a fourth order boundary-value problem
\begin{equation}
    \frac{{\rm d}^4 Z_0}{{\rm d} X^4}+\frac{1}{\beta}=0,
    \label{eq:linearized_beam_statics}
\end{equation}
 with boundary conditions ${\rm d}^2Z_0/{\rm d}^2X(X=0,1)=\kappa_0/|\kappa_0|$ and $Z_0(X=0,1)=0$;  the latching constraint is $Z_0(X=1/2)=0$, but the force required to achieve this, $F_0$, is at present unknown. We have seen in~\S~\ref{sec:latch} how the latch constraint induces a discontinuity in the third derivative, $Z_0$. As noted above, we split the domain of the $X$ coordinates into two: the region to the left of the latch, $X_L\in[0,1/2]$, and the region to the right, $X_R\in[1/2,1]$, such that the discontinuity is now at the boundary of each subdomain, and hence we can find a well-behaved function that solves \Eq{\ref{eq:linearized_beam_statics}} on each side of the latch.
The solution of this system gives the following polynomial for $Z_0(X)$:
\begin{align}
    Z&_0(X)=\nonumber\\
    &\begin{cases}
        -\frac{1}{24\beta}X^4- \frac{1}{2}\left(\frac{\kappa_0}{|\kappa_0|}-\frac{1}{16\beta}\right)X^3+\frac{1}{2}\frac{\kappa_0}{|\kappa_0|}X^2-\frac{1}{8}\left(\frac{1}{48}\frac{1}{\beta}+\frac{\kappa_0}{|\kappa_0|}\right)X,\quad\text{for }X\in[0,1/2]\\
        -\frac{1}{24\beta}\hat X^4+ \frac{1}{2}\left(\frac{\kappa_0}{|\kappa_0|}-\frac{1}{16\beta}\right)\hat X^3+\frac{1}{2}\frac{\kappa_0}{|\kappa_0|}\hat X^2+\frac{1}{8}\left(\frac{1}{48}\frac{1}{\beta}+\frac{\kappa_0}{|\kappa_0|}\right)\hat X, \quad\text{for }\hat X=X-1,\,X\in[1/2,1],
    \end{cases}
    \label{eq:Z0}
\end{align}
where we have written the solution to the right of the latch as a function of a transformed coordinate $\hat X$ that exploits the symmetry of the shell with respect to its center, $X=1/2$.

\subsubsection*{\label{sec:nonlinear}Non-linear model}
We complete this section with an analytic solution to the equilibrium, fully non-linear model \Eqs{\ref{eq:balance_angular_momentum_sc}, \ref{eq:balance_force_x_sc} \& \ref{eq:balance_force_z_sc}} with boundary conditions by \Eq{\ref{eq:bcs_sc}}.
Substituting $\Lambda_X=0$ (this result follows from the boundary conditions, $\left.\Lambda_X(S)\right|_{S=0,1}=0$, combined with the equation $\frac{{\rm d}\Lambda_X}{{\rm d}S}=0$, \Eq{\ref{eq:balance_force_x_sc}} in equilibrium) into the equilibrium \Eq{\ref{eq:balance_angular_momentum_sc}}
\begin{equation}
    \frac{{\rm d}^2\theta}{{\rm d}S^2}-\frac{\rho g h \ell^3}{EI}(1-S)\cos{\theta}+\ell|\kappa_0|\Lambda_Z\cos\theta-\ell|\kappa_0|\Lambda_g\cos\theta=0,
\label{eq:nonlinear}
\end{equation}
where $\Lambda_Z$ is a constant, yet unknown, by virtue of \Eq{\eqref{eq:balance_force_z_sc}} in equilibrium.
This equation is a non-linear and non-autonomous differential equation that is not solvable analytically for arbitrary values of the parameters $\rho g h\ell^3/EI$ and $\ell|\kappa_0|$ (cf.~the equation for a hair strand in~\cite{AudolyPomeau2010}). Yet it can be solved in the limit of weak gravity (i.e.~very stiff, very light and/or very short shell) for which we do not need a force (i.e.~the Lagrange multiplier $\Lambda_g$) to constrain the position of the edges, and the equation~\ref{eq:nonlinear} becomes
\begin{equation}
    \frac{{\rm d}^2\theta}{{\rm d}S^2}+\ell|\kappa_0|\Lambda_Z\cos\theta=0,
\label{eq:nonlinear_weakG}
\end{equation} with moment-free boundary conditions at $S=0,1$:
\begin{equation}
\left.   \frac{{\rm d}\theta}{{\rm d}S}\right|_{S=0,1}=\ell\kappa_0.
\label{eq:bcs_nolin}
\end{equation}
Invoking the same arguments as in \S\ref{sec:latch}, we notice that $\Lambda_Z$ accounts for the vertical forces by the latch mechanism and the corresponding reaction from the substrate. Therefore, by integrating \Eq{\ref{eq:nonlinear_weakG}} along the two subdomains $S_L\in[0,1/2]$ and $S_R\in[1/2,1]$ we obtain
\begin{eqnarray}
    \Lambda_{Z}^{L}(S=0)=-R_0,\quad \Lambda_{Z}^{L}(S=1/2)=\frac{1}{2}F_0,\nonumber\\ \Lambda_{Z}^{R}(S=1/2)=-\frac{1}{2}F_0,\quad
    \text{and}\quad \Lambda_{Z}^{R}(S=1)=R_0,
    \label{eq:forces_nolin}
\end{eqnarray}
and hence $\theta(S)$ exhibits a discontinuity in the second derivative due to the latch force, i.e. combining \Eqs{\ref{eq:nonlinear_weakG} \& \ref{eq:forces_nolin}} we get:
\begin{equation}
    \left[\frac{{\rm d}^2\theta}{{\rm d}S^2}\right]_{S=1/2^-}^{S=1/2^+}=F_0.
    \label{eq:force_nolin}
\end{equation}
$\Lambda_Z$ (and hence $F_0$) on each side of the latch are given now by the non-linear constraints
\begin{equation}
    \int_0^{1/2}\sin\theta(S)\,\mathrm{d}S=0,\quad{\rm and}\quad \int_{1/2}^{1}\sin\theta(S)\,\mathrm{d}S=0.
    \label{eq:nolin_constraints}
\end{equation}

The integral of \Eq{\ref{eq:nonlinear_weakG}} can be expressed in elliptic-function form~\cite{AudolyPomeau2010}:
\begin{align}
    \theta_{L}^{\pm}(S, \Lambda_L, \theta_0, \ell, \kappa_0) = \frac{\pi}{2} - 2 \, \text{am} \left({\mathcal F}\left( \frac{\pi - 2\theta_0}{4}, k_L \right)\pm \frac{1}{2} \sqrt{\Delta_L} \, S,\ k_L \right)\quad{\rm for}\,S\in[0,1/2]\label{eq:sols_nonlinear_L}\\
    \theta_{R}^{\pm}(S, \Lambda_R, \theta_1, \ell, \kappa_0) = \frac{\pi}{2} - 2 \, \text{am} \left({\mathcal F}\left( \frac{\pi - 2\theta_1}{4}, k_R \right)\pm \frac{1}{2} \sqrt{\Delta_R} \, (S - 1),\ k_R \right)\quad{\rm for}\,S\in[1/2,1],
    \label{eq:sols_nonlinear_R}
\end{align}
where $\text{am}(\cdot,k)$ is the Jacobi amplitude function, ${\mathcal F}(\phi,k)$ is the incomplete elliptic integral of the first kind, $\theta_{(0,1)}$ is the angle at the edges, and
\begin{align*}
k_L &= \frac{-4 \ell |\kappa_0| \Lambda_Z^L}{(\kappa_0 \ell)^2 - 2\ell |\kappa_0| \Lambda_Z^L + 2\ell |\kappa_0| \Lambda_Z^L \sin(\theta_0)}, \\
k_R &= \frac{-4 \ell |\kappa_0| \Lambda_Z^R}{(\kappa_0 \ell)^2 - 2\ell |\kappa_0| \Lambda_Z^R + 2\ell |\kappa_0| \Lambda_Z^R \sin(\theta_1)}, \\
\Delta_L &= (\kappa_0 \ell)^2 - 2\ell |\kappa_0| \Lambda_Z^L + 2\ell |\kappa_0| \Lambda_Z^L \sin(\theta_0), \\
\Delta_R &= (\kappa_0 \ell)^2 - 2\ell |\kappa_0| \Lambda_Z^R + 2\ell |\kappa_0| \Lambda_Z^R \sin(\theta_1).
\end{align*}

We have 4 solutions \Eqs{\ref{eq:sols_nonlinear_L} \& \ref{eq:sols_nonlinear_R}} containing 4 unknown constants ($\theta_0,\theta_1,\Lambda_Z^L,\Lambda_Z^R$) that we will determine by satisfying the  boundary conditions \Eq{\ref{eq:bcs_nolin}} and the constraints \Eq{\ref{eq:nolin_constraints}}.

To integrate the constraints in \Eq{\ref{eq:nolin_constraints}} we follow the same steps as in~\cite{WagnerVella13}: considering the example of $\theta_L$, we first change variables
\begin{align*}
    \int_0^{1/2}{\rm d}S\,\sin{\theta_L(S)}=\int_{\theta_0}^{\theta_{1/2}}{\rm d}\theta_L\left(\frac{{\rm d}\theta_L}{{\rm d}S}\right)^{-1}\,\sin{\theta_L},
\end{align*}
for which \[\frac{{\rm d}\theta_L}{{\rm d}S}=\pm\left[2\left(\frac{1}{2}+\ell|\kappa_0|\Lambda_Z^{L}\sin{\theta_0}-\ell|\kappa_0|\Lambda_Z^{L}\sin{\theta_L}\right)\right]^{1/2}\] is defined up to a sign.
We then split the integral into two 
\begin{align*}
    \int_{\theta_0}^{\theta_{1/2}}{\rm d}\theta_L\left(\frac{{\rm d}\theta_L}{{\rm d}S}\right)^{-1}\,\sin{\theta_L}=-\int_{\theta_0}^{\theta_{S}}{\rm d}\theta_L\left(\frac{{\rm d}\theta_L}{{\rm d}S}\right)^{-1}\,\sin{\theta_L}+\int_{\theta_S}^{\theta_{1/2}}{\rm d}\theta_L\left(\frac{{\rm d}\theta_L}{{\rm d}S}\right)^{-1}\,\sin{\theta_L},
\end{align*}
where $\theta_S$ is the angle at $S=S_0$, with $S_0$ such that $\left.\frac{{\rm d}\theta_L}{{\rm d}S}\right|_{S=S_0}=0$, and we have introduced the sign of the first derivative explicitly. The integral constraints \Eq{\ref{eq:nolin_constraints}} are then
\begin{align}
{\rm C}_L&(\theta_0, S_{0}, \Lambda_Z^L, \ell, \kappa_0) 
=\nonumber\\
&\frac{1}{\ell |\kappa_0| \Lambda_Z^L} \Bigg[ 
\sqrt{\Delta_L} \left( 
    {\cal E}\left( \tfrac{1}{4}(\pi - 2\theta_0), k_L \right)
    - 2 {\cal E}\left( \tfrac{1}{4}(\pi - 2\theta_{L}^+(S_{0})), k_L \right)
    + {\cal E}\left( \tfrac{1}{4}(\pi - 2\theta_{L}^+(1/2)), k_L \right)
\right) \nonumber\\
& - \frac{(\kappa_0 \ell)^2 + 2\ell |\kappa_0| \Lambda_Z^L \sin(\theta_0)}{\sqrt{\Delta_L}} \left(
    {\cal F}\left( \tfrac{1}{4}(\pi - 2\theta_0), k_L \right)
    - 2 {\cal F}\left( \tfrac{1}{4}(\pi - 2\theta_{L}^+(S_{0})), k_L \right)\nonumber\right.\\
&    \left.+ {\cal F}\left( \tfrac{1}{4}(\pi - 2\theta_{L}^+(1/2)), k_L \right)
\right)
\Bigg],\quad{\rm and}\\
{\rm C}_R&(\theta_1, S_{0}, \Lambda_Z^R, \ell, \kappa_0) 
=\nonumber\\
&\frac{1}{\ell |\kappa_0| \Lambda_Z^R} \Bigg[ 
\sqrt{\Delta_R} \left( 
    -{\cal E}\left( \tfrac{1}{4}(\pi - 2\theta_1), k_R \right)
    + 2 {\cal E}\left( \tfrac{1}{4}(\pi - 2\theta_{R}^-(S_{0})), k_R \right)
    - {\cal E}\left( \tfrac{1}{4}(\pi - 2\theta_{R}^-(1/2)), k_R \right)
\right) \nonumber\\
& - \frac{(\kappa_0 \ell)^2 + 2\ell |\kappa_0| \Lambda_Z^R \sin(\theta_1)}{\sqrt{\Delta_R}} \left(
    -{\cal F}\left( \tfrac{1}{4}(\pi - 2\theta_1), k_R \right)
    + 2 {\cal F}\left( \tfrac{1}{4}(\pi - 2\theta_{R}^-(S_{0})), k_R \right)\nonumber\right.\\
&    \left.- {\cal F}\left( \tfrac{1}{4}(\pi - 2\theta_{R}^-(1/2)), k_R \right)
\right)
\Bigg],
\label{eq:int_constraints}
\end{align}
where ${\mathcal E}(\phi,k)$ is the incomplete elliptic integral of the second kind.

Finally, we compute numerically the six unknowns ($\theta_0,\theta_1,\Lambda_Z^L,\Lambda_Z^R,S_0^L,S_0^R$) from the system of six equations:
\begin{eqnarray}
\begin{cases}
    \left.\left(\theta_R^--\theta_L^+\right)\right|_{S=1/2}=0,\\
    \left.\left(\frac{{\rm d}\theta_R^+}{{\rm d}S}-\frac{{\rm d}\theta_L^+}{{\rm d}S}\right)\right|_{S=1/2}=0,\\
    {\rm C}_L(\theta_0, S_{0}^L, \Lambda_Z^L, \ell, \kappa_0) = 0\\
    {\rm C}_R(\theta_1, S_{0}^R, \Lambda_Z^R, \ell, \kappa_0) = 0\\
    \left.\frac{{\rm d}\theta_L^+}{{\rm d}S}\right|_{S=S_0^L}=0,\\
    \left.\frac{{\rm d}\theta_R^+}{{\rm d}S}\right|_{S=S_0^R}=0.
\end{cases}
\label{eq:system_nolin}
\end{eqnarray}
In \Fig{3} of the main text we plot the latch force $\Lambda_Z^R-\Lambda_Z^L$ (cf. \Eq{\ref{eq:force_nolin}}) using the values for $\Lambda_Z^L,\Lambda_Z^R$  obtained by solving this system numerically.

\subsection*{\label{sec:release}Dynamic beam equation: the latch is released}
The dynamic beam equation~\ref{eq:linearized_beam_damped} with the boundary, initial and continuity conditions discussed above can be solved analytically by letting
\begin{equation}
    Z(X,T)=\zeta(X,T)-w(X)-\exp(-T/\tau)\,y(X),
    \label{eq:Z}
\end{equation}
where $\zeta(X,T)$ is the separable solution of the homogeneous problem, $w(X)$ is a particular solution that absorbs the inhomogeneous parts arising from the natural curvature, and $y(X)$ is a particular solution that absorbs the discontinuity imposed by \Eq{\ref{eq:discont_3rd_deriv_dyn}}. Specifically, we find that
 \begin{equation}
     w(X)=\frac{1}{2}\Bigg[\frac{1}{12}\frac{1}{\beta}X^4-\frac{1}{6}\frac{1}{\beta}X^3-\frac{\kappa_0}{|\kappa_0|}\,X^2+\left(\frac{1}{12}\frac{1}{\beta}+\frac{\kappa_0}{|\kappa_0|}\right)X\Bigg],
     \label{eq:w}
 \end{equation} while
\begin{align}
    y(x)=
    \begin{cases}
    y_L(X),\quad \text{for} \quad X\in[0,1/2]\\
    -y_L(\hat X),\,\hat X=X-1,\quad \text{for} \quad X\in[1/2,1];
    \end{cases}
    \label{eq:y}
\end{align}
 with
 \begin{align}
    y_L(X)=&\left[\frac{{\rm d}^3 Z_0(X)}{{\rm d} X^3}\right]_{X=1/2^-}^{X=1/2^+}\left(\frac{\tau}{2}\right)^{3/2}\frac{1}{(1-H\tau)^{3/4}}\Bigg\{ c_1^L\sinh\left[\frac{(1-H\tau)^{1/4}}{\sqrt{2 \tau}}X\right]\cos\left[\frac{(1-H\tau)^{1/4}}{\sqrt{2 \tau}}X\right]\nonumber\\
    &+c_2^L\cosh\left[\frac{(1-H\tau)^{1/4}}{\sqrt{2 \tau}}X\right]\sin\left[\frac{(1-H\tau)^{1/4}}{\sqrt{2 \tau}}X\right]\Bigg\},
 \label{eq:yL}
 \end{align} for $H\tau<1$ (i.e.~rapid release compared to the intrinsic damping scale, $1/H$).  Here, $c_1^L,c_2^L$ are the constants:
 \begin{eqnarray}
    c_1^L&=-\frac{\cosh{\frac{(1-H\tau)^{1/4}}{2\sqrt{2\tau}}}\cos{\frac{(1-H\tau)^{1/4}}{2\sqrt{2\tau}}}+\sinh{\frac{(1-H\tau)^{1/4}}{2\sqrt{2\tau}}}\sin{\frac{(1-H\tau)^{1/4}}{2\sqrt{2\tau}}}}{\cosh^2{\frac{(1-H\tau)^{1/4}}{2\sqrt{2\tau}}}\cos^2{\frac{(1-H\tau)^{1/4}}{2\sqrt{2\tau}}}+\sinh^2{\frac{(1-H\tau)^{1/4}}{2\sqrt{2\tau}}}\sin{\frac{(1-H\tau)^{1/4}}{2\sqrt{2\tau}}}},\\
    c_2^L&=\frac{\cosh{\frac{(1-H\tau)^{1/4}}{2\sqrt{2\tau}}}\cos{\frac{(1-H\tau)^{1/4}}{2\sqrt{2\tau}}}-\sinh{\frac{(1-H\tau)^{1/4}}{2\sqrt{2\tau}}}\sin{\frac{(1-H\tau)^{1/4}}{2\sqrt{2\tau}}}}{\cosh^2{\frac{(1-H\tau)^{1/4}}{2\sqrt{2\tau}}}\cos^2{\frac{(1-H\tau)^{1/4}}{2\sqrt{2\tau}}}+\sinh^2{\frac{(1-H\tau)^{1/4}}{2\sqrt{2\tau}}}\sin{\frac{(1-H\tau)^{1/4}}{2\sqrt{2\tau}}}}.
 \end{eqnarray}

For completeness, we note that the separable solution is:
 \begin{equation}
 \zeta(X,T)=e^{-\tfrac{H}{2}T}\sum_{n=1}^{\infty}\sin\left(n\pi X\right)A_n\cos(\omega_n T+\varphi_n),\label{eq:zeta}
 \end{equation}
where $\omega_n=\frac{H}{2}\left(\frac{4 n^4\pi^4}{H^2}-1\right)^{1/2}$ are  the frequencies of natural oscillations  in underdamped conditions ($H<2 n^2\pi^2$).  The $\varphi_n$ are phase differences determined from the initial conditions, which are now written as
$\zeta(X,0)=Z_0(X)+w(X)+y(X)$ and, using \Eq{\ref{eqn:ZdotIC}}, we have $\left.\frac{\partial\zeta}{\partial T}\right|_{T=0}=-\frac{y(X)}{\tau}.$

The constants $A_n$ and $\varphi_n$ are
\begin{equation}
    A_n=\sign{\left(a_n\right)}\sqrt{a_n^2+b_n^2}\quad \text{and}\quad \varphi_n=\arctan{\left(\frac{-b_n}{a_n}\right)},
    \label{eq:Avarphi}
\end{equation}
where $a_n,b_n$ are the coefficients of the Fourier sine series of the relevant initial conditions, i.e.
\begin{align}
    a_n=&2\int_0^1{\rm d}X\,\zeta(X,0)\sin{n\pi X}\nonumber\\
    =&2\left[\frac{{\rm d}^3 Z_0(X)}{{\rm d} X^3}\right]_{X=1/2^-}^{X=1/2^+}\frac{1}{1+\frac{n^4\pi^4\tau^2}{1-H\tau}}\frac{1}{n^4\pi^4}\sin{n\frac{\pi}{2}},\label{eq:an}\\
    b_n=&\frac{2}{(n\pi)^2}\int_0^1{\rm d}X\,\frac{\partial\zeta}{\partial T}(X,0)\sin{n\pi X}\nonumber\\
    =&\frac{1}{\omega_n}\left[\frac{{\rm d}^3 Z_0(X)}{{\rm d} X^3}\right]_{X=1/2^-}^{X=1/2^+}\frac{1}{1+\frac{n^4\pi^4\tau^2}{1-H\tau}}\frac{1}{n^4\pi^4}\sin{n\frac{\pi}{2}}\left[\frac{H}{2}+\frac{n^4\pi^4\tau}{1-H\tau}\right].\label{eq:bn}
\end{align}

\subsubsection*{\label{sec:overdamped}Slow release ($H\tau\geq 1$)}
We include for completeness the analytic results of the dynamic beam equation~\ref{eq:linearized_beam_damped} when the unlatching timescale $\tau$ is larger than the damping timescale $H^{-1}$ --- yet $H< 2\pi^2$ and hence all modes are oscillating. For $H\tau>1$, the solution is given by \Eq{\ref{eq:Z}} with $w(X)$ from \Eq{\ref{eq:w}} and $\zeta(X,T)$ from \Eq{\ref{eq:zeta}} (with the same constants as in \Eqs{\ref{eq:Avarphi},~\ref{eq:an} \&~\ref{eq:bn}}. $y(X)$ is as it is defined in \Eq{\ref{eq:y}} with $y_L(X)$:
 \begin{align}
    y_L(X)=\left[\frac{{\rm d}^3 Z_0(X)}{{\rm d} X^3}\right]_{X=1/2^-}^{X=1/2^+}\frac{\tau^{3/2}}{4}\frac{1}{(H\tau-1)^{3/4}}\left\{\frac{\sinh\left[\frac{(H\tau-1)^{1/4}}{\sqrt{\tau}}X\right]}
    {\cosh\left[\frac{(H\tau-1)^{1/4}}{2\sqrt{\tau}}\right]}
    -\frac{\sin\left[\frac{(H\tau-1)^{1/4}}{\sqrt{\tau}}X\right]}
    {\cos\left[\frac{(H\tau-1)^{1/4}}{2\sqrt{\tau}}\right]}\right\}.
 \label{eq:yLg1}
 \end{align}

 For $H\tau=1$ the solution is also constructed as in \Eq{\ref{eq:Z}}, with $w(X)$ given by \Eq{\ref{eq:w}}, $y(X)$ by \Eq{\ref{eq:y}}, where $y_L(X)$ is
 \begin{equation}
    y_L(X)=\frac{1}{4}\left[\frac{{\rm d}^3 Z_0(X)}{{\rm d} X^3}\right]_{X=1/2^-}^{X=1/2^+}\left(\frac{1}{3}X^3-\frac{1}{4}X\right),
    \label{eq:yL1}
 \end{equation}
 and the separable solution $\zeta(X,T)$ is given by
 \begin{equation}
    \zeta(X,T)=4\left[\frac{{\rm d}^3 Z_0(X)}{{\rm d} X^3}\right]_{X=1/2^-}^{X=1/2^+}e^{-\tfrac{T}{2\tau}}\sum_{n=1}^{\infty}\frac{\sin{n\frac{\pi}{2}}}{(n\pi)^4(4\tau^2n^4\pi^4-1)^{1/2}}\sin\left(n\pi X\right)\sin(\omega_n T),\label{eq:zeta1}
 \end{equation}
with the frequencies of natural oscillations now given by $\omega_n=\frac{1}{2\tau}\left(4 n^4\pi^4\tau^2-1\right)^{1/2}$.

\section{\label{sec:jumping}Jumping behavior}
\subsection*{\label{sec:reaction}Reaction force at $X=0$}
We analyze  the jumping behavior in terms of the three dimensionless parameters in the model: $\beta,\tau$ and $H$.
We begin by noting that the scaled reaction force of the substrate on the edge of the shell, $R(T)$, may be calculated from the solution $Z(X,T)$, \Eq{\ref{eq:Z}}, as
\begin{equation}
    R(T)=\left.\frac{\partial^3Z}{\partial X^3}\right|_{X=0}.
\end{equation}
This result follows from evaluating \Eq{\ref{eq:balance_forces}} at $X=0$ and substituting the result $\Lambda_Z(0,T)=-R(T)$, obtained from the integral of \Eq{\ref{eq:equilibrium_release}}.

The reaction force can only be positive --- if it becomes negative then the shell has lost contact with the ground, and the shell has jumped.
Finding the smallest positive root of $R(T)=0$ therefore determines the jumping time, $T_{\rm jump}$.
The analytic expressions of the force at $X=0$ for the different solutions in \S~\ref{sec:solutions} are:
\begin{align}
    R(0,T)=
    \begin{cases}
    \frac{1}{2\beta}-\left[\frac{{\rm d}^3 Z_0(X)}{{\rm d} X^3}\right]_{X=1/2^-}^{X=1/2^+}\left\{\sum_{n=1}^{\infty}\frac{2}{1+\frac{n^4\pi^4\tau^2}{1-H\tau}}\frac{1}{n\pi}\sin{\left(n\frac{\pi}{2}\right)}e^{-\tfrac{H}{2}T}\left[\cos{\omega_n T}+\frac{1}{\omega_n}\left(\frac{H}{2}+\frac{n^4\pi^4\tau}{1-H\tau}\right)\sin{\omega_n T}\right]\right.\\
    \quad\left.+\frac{1}{2}e^{-\tfrac{T}{\tau}}\frac{\cosh{\frac{(1-H\tau)^{1/4}}{2\sqrt{2\tau}}}\cos{\frac{(1-H\tau)^{1/4}}{2\sqrt{2\tau}}}}{\cosh^2{\frac{(1-H\tau)^{1/4}}{2\sqrt{2\tau}}}\cos^2{\frac{(1-H\tau)^{1/4}}{2\sqrt{2\tau}}}+\sinh^2{\frac{(1-H\tau)^{1/4}}{2\sqrt{2\tau}}}\sin^2{\frac{(1-H\tau)^{1/4}}{2\sqrt{2\tau}}}}\right\},\quad\text{for }H\tau<1;\\
    \frac{1}{2\beta}-\left[\frac{{\rm d}^3 Z_0(X)}{{\rm d} X^3}\right]_{X=1/2^-}^{X=1/2^+}\left[\frac{1}{2}e^{-\tfrac{T}{\tau}}+4e^{-\tfrac{T}{2\tau}}\sum_{n=1}^{\infty}\frac{\sin{n\frac{\pi}{2}}}{n\pi(4\tau^2n^4\pi^4-1)^{1/2}}\sin{\omega_n T}\right],\quad\text{for }H\tau=1;\\
    \frac{1}{2\beta}-\left[\frac{{\rm d}^3 Z_0(X)}{{\rm d} X^3}\right]_{X=1/2^-}^{X=1/2^+}\left\{e^{-\frac{H}{2}T}\sum_{n=1}^{\infty}\frac{4\sin{n\frac{\pi}{2}}}{n\pi(4\tau^2n^4\pi^4-1)^{1/2}}\sin{\omega_n T}\right.\\
    \quad\left.+e^{\tfrac{-T}{\tau}}\frac{1}{4}\left[\frac{1}{\cosh{\frac{(H\tau-1)^{1/4}}{2\sqrt{\tau}}}}+\frac{1}{\cos{\frac{(H\tau-1)^{1/4}}{2\sqrt{\tau}}}}\right]\right\},\quad\text{for }H\tau>1.
    \end{cases}
    \label{eq:R}
\end{align}
Recall that the natural frequencies are $\omega_n=\frac{H}{2}\left(\frac{4 n^4\pi^4}{H^2}-1\right)^{1/2}$ for $H\tau\neq1$, and $\omega_n=\frac{1}{2\tau}\left(4 n^4\pi^4\tau^2-1\right)^{1/2}$ for $H\tau=1$.

\subsection*{\label{sec:phases}Phase diagram}
\begin{figure}[htbp]
  \centering
     \begin{overpic}[width=0.8\linewidth]{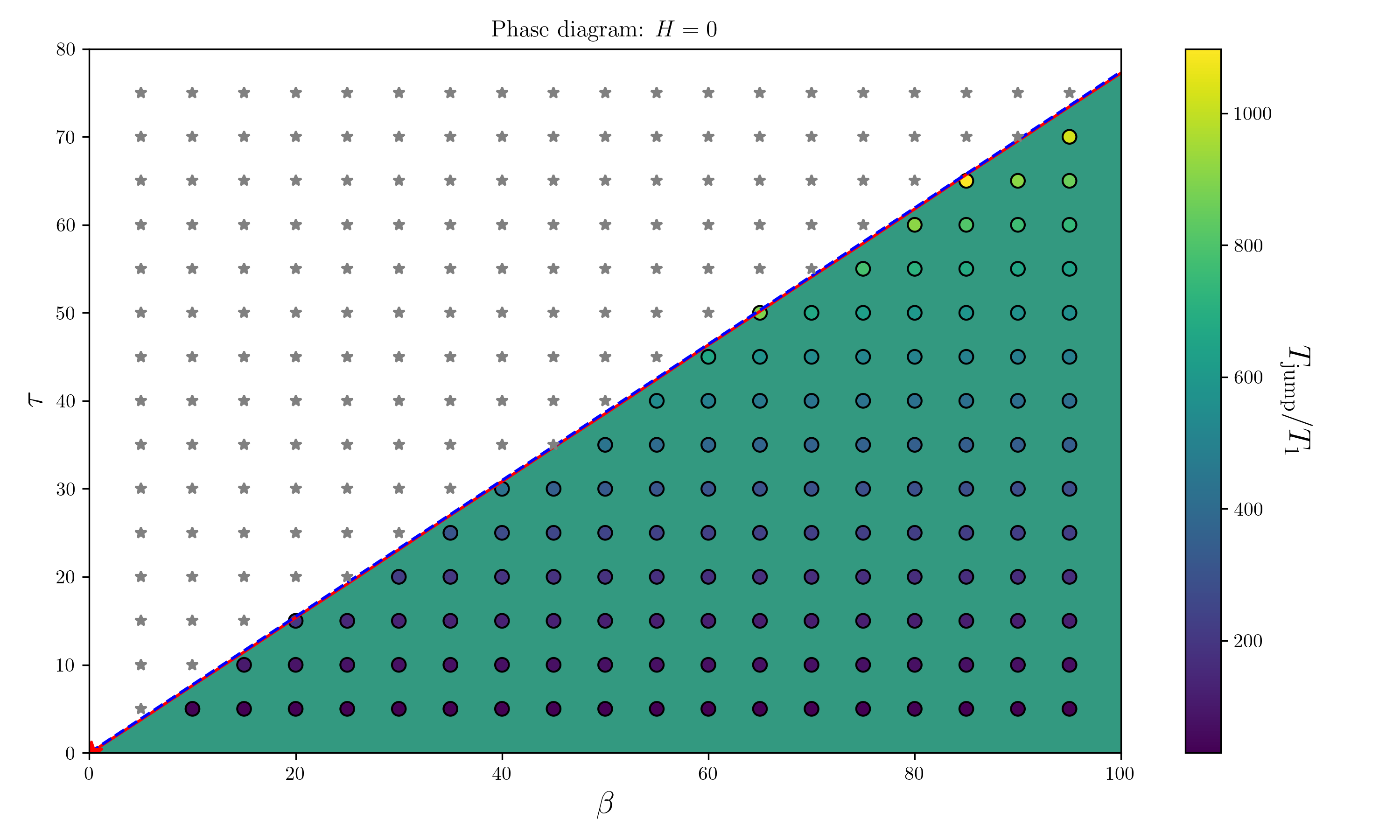}\put(0,55){\textbf{A}}\end{overpic}
     \begin{overpic}[width=0.8\linewidth]{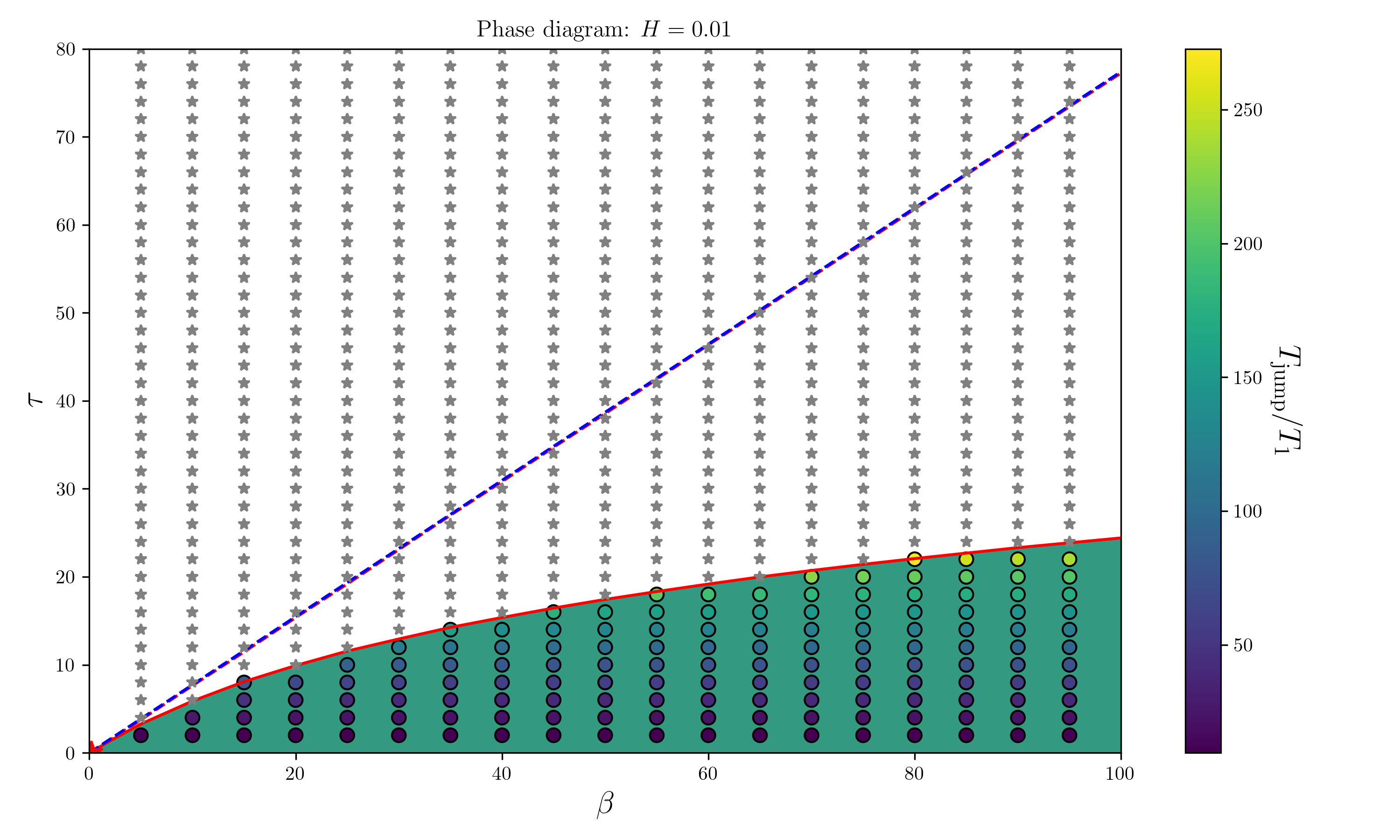}\put(0,55){\textbf{B}}\end{overpic}
\caption{\label{fig:phase} Phase diagram of the jumping behavior of a curved shell on a flat surface at different values of the damping $H=0$ (\textbf{A}), and $H=0.01$ (\textbf{B}). The jumping behavior of the shell is characterized by the triad $(\beta,\tau,H)$: there is a jumping phase (shaded in green) and a non-jumping phase (shaded in white). Grey stars indicate failed attempts and dots indicate jumps with jumping time $T_{\rm jump}$ given by the color bar to the right. The asymptotic expression for the phase boundary \Eq{\ref{eq:tauboundary}} is plotted with a blue dashed line.}
\end{figure}

We sample a grid of points $(\beta,\tau,H)$ in the phase space and find for each of them the roots of $R(T)$ in \Eq{\ref{eq:R}} numerically with the \emph{Chebfun} package~\cite{Chebfun} in \emph{Matlab}. For this computation the series \Eq{\ref{eq:R}} is truncated at mode $n_{\rm modes}=\left\lceil \sqrt{2/(\pi\tau)} \right\rceil$, given by the condition that the smallest time that we resolve compares with $\tau$, i.e. $\left(\omega_{n_{\rm modes}}\right)^{-1}\sim\tau$. This choice can be justified with the physical argument that higher modes should be suppressed by the external force due to the latch-release mechanism. We record the time of jumping $T_{\rm jump}$ and normalize it with the period of the fundamental mode $T_1=2\pi/\omega_1$. We show the results for $H=0$ and $H=0.01$ in \Fig{\ref{fig:phase}}.

\subsubsection*{\label{sec:asymptotics}Asymptotic expressions of the phase boundary} 
Having computed the phase boundaries in \Fig{\ref{fig:phase}} numerically, we now seek an asymptotic approximation of the force at the origin \Eq{\ref{eq:R}} that will allow us to determine asymptotic results for the phase boundary and $\beta_{\rm crit}$, which is the minimal $\beta$ needed for jumping.

We start by considering the asymptotic limit of $H\ll 1$ (undamped), slow release $\tau\gg 1$ and large time $T\gg\tau$ for a light shell ($\beta\gg1$) in $R(0,T)$ with $H\tau<1$ using \Eq{\ref{eq:R}}:
\begin{equation}
    R(0,T)\approx \frac{1}{2\beta}+6\sum_{n=1}^{\infty}\frac{2\sin{(n^2\pi^2 T)}}{n^3\pi^3\tau}\sin{\left(n\frac{\pi}{2}\right)}.
    \label{eq:Rasym1}
\end{equation}
Here we have used the asymptotic limit of $\beta\gg 1$ for the discontinuity in the third derivative of the statics $Z_0(X)$ from \Eq{\ref{eq:Z0}}.
This limit accounts for the largest $\tau$ (slowest release) for which jumping occurs --- considering otherwise the most favorable conditions for jumping for $\beta$ and $H$, i.e.~a light object with no damping. The condition for no jumping corresponds to when the series attains a local minimum and $R(0,T)=0$. A good estimate of this behavior is given by the local minimum of the main mode $(n=1)$, and hence we assume $\sin{(\pi^2T)}=-1$ in \Eq{\ref{eq:Rasym1}}. The condition $R(0,T)=0$ then poses a linear equation for $\tau$ that we can solve, and we obtain
\begin{equation}
    \tau_{\rm boundary}(\beta)=\frac{24}{\pi^3}\beta.
    \label{eq:tauboundary}
\end{equation}

We now examine the opposite limit of quick release $\tau\ll 1$ and heavy shell (i.e. small $\beta$); the asymptotic expression for $R(0,T)$ \Eq{\ref{eq:R}} is, assuming again a large time $T\gg\tau$ and $H\ll 1$, given by
\begin{equation}
    R(0,T,n_{\rm modes})\approx \frac{1}{2\beta}-\left[\frac{{\rm d}^3 Z_0(X)}{{\rm d} X^3}\right]_{X=1/2^-}^{X=1/2^+}\sum_{n=1}^{n_{\rm modes}}\frac{2\cos{(n^2\pi^2 T)}}{n\pi}\sin{\left(n\frac{\pi}{2}\right)},
    \label{eq:Rasym2}
\end{equation}
where we have assumed $n_{\rm modes}^4\pi^4\tau^2\ll 1$.
We will further approximate $R(0,T,n_{\rm modes})$ by truncating the series at the main mode $(n=1)$. The condition of no jump is then given by
\begin{equation}
R(0,T,1)-\max_T|R(0,T,n_{\rm modes})-R(0,T,1)|=0,
\label{eq:cond_nojump}
\end{equation}
where we have subtracted the maximum amplitude of the residual series after removing the fundamental mode. To estimate this amplitude, we will use the fact that it is bounded by the root mean square (RMS) amplitude of the residual series.
Given that the series in \Eq{\ref{eq:Rasym2}} converges and the basis functions $\cos{(n^2\pi^2 T)}$ are bounded and quasi-orthogonal, we can calculate the RMS amplitude
\begin{equation}
    {\rm RMS}^2[R(0,T,n_{\rm modes})-R(0,T,1)]\approx \frac{1}{2}\left\{\frac{2}{\pi}\left[\frac{{\rm d}^3 Z_0(X)}{{\rm d} X^3}\right]_{X=1/2^-}^{X=1/2^+}\right\}^2\sum_{n=3,\,n~{\rm odd}}^{{\infty}}\frac{1}{n^2}.
\end{equation}
Applying the well known result \[\sum_{n=1,\,n~{\rm odd}}^{\infty}\frac{1}{n^2}=\frac{\pi^2}{8},\] we obtain
\begin{align}
    {\rm RMS}^2[R(0,T,n_{\rm modes})-R(0,T,1)]\approx \frac{1}{2}\left\{\frac{2}{\pi}\left[\frac{{\rm d}^3 Z_0(X)}{{\rm d} X^3}\right]_{X=1/2^-}^{X=1/2^+}\right\}^2\left(\frac{\pi^2}{8}-1\right).
\end{align}
We estimate from our numerical solutions that the critical value for jumping is $\beta\approx 0.2$ (see \Fig{2b} in the main text). We then estimate the discontinuity in the third derivative for this value of $\beta$ (using \Eq{\ref{eq:Z0}}) and find that ${\rm RMS}[R(0,T,n_{\rm modes})-R(0,T,1)]\sim 1$; thus can we rewrite the condition of no jump \Eq{\ref{eq:cond_nojump}} as
\begin{align}
    R(0,T,1)-1=0.
\end{align}
We solve this equation and obtain
\begin{equation}
    \beta_{\rm crit}=\frac{5+2\pi}{4(12+\pi)}\approx 0.186.
    \label{eq:betacrit}
\end{equation}
We plot this estimate for $\beta_{\rm crit}$ in fig.~2b in the main text, and note that the approximate value in \Eq{\ref{eq:betacrit}} is consistent with our numerical results obtained without this truncation.

A rougher estimate of $\beta_{\rm crit}$ is given by considering a sufficiently heavy shell that bends due to its own weight and hence naturally satisfies the condition $Z_0(X=1/2)=0$ at zero external force. Hence we impose the continuity in the third derivative
\begin{equation}
    \left[\frac{{\rm d}^3 Z_0}{{\rm d}X^3}\right]_{X=1/2^-}^{X=1/2^+}=0,
    \label{eq:cond_betacrit}
\end{equation}
and solving the equation \[\frac{\rm{d}^3Z_0}{\rm{d}X^3}\Bigg|_{X=1/2}=0\] we obtain
\begin{equation}
\beta_{\rm{crit}}=\frac{5}{48}\approx 0.104.
\end{equation}

\subsection*{\label{sec:efficiency}Efficiency of jumping}

We define the efficiency of jumping as the ratio of translational kinetic energy at the point of jumping, $K_{\rm trans}$, to the stored  energy $E_0$. Here, $E_0$ is the sum of the bending and gravitational potential energy of the latched shell, i.e.~for the solution in \Eq{\ref{eq:Z0}},
\begin{equation}
    E_0(\beta)=\frac{1}{2}\left[\int_0^{1}{\rm d}X \left(\frac{{\rm d}^2Z_0}{{\rm d}X^2}-\frac{\kappa_0}{|\kappa_0|}\right)^2\right]+\frac{1}{\beta}\int_0^1{\rm d}X\,Z_0=\frac{-3+160\beta(1+72\beta)}{30720\beta^2};
    \label{eq:stored_energy}
\end{equation}
the translational kinetic energy is
\begin{equation}
    K_{\rm trans}=\frac{1}{2}\left[\frac{\partial}{\partial T}\int_0^1{\rm d}X\, Z(X,T)\right]^2,
\end{equation} directly from the solution in \Eq{\ref{eq:Z}}.

In \Fig{4} of the main text there are certain \emph{pessimal} values of $\tau$ for which the efficiency drops dramatically. We notice that these pessimal values correspond to when $\tau$ is close to those jumping times $T_{\rm jump}$ such that $R(T_{\rm jump})$ is  a local minimum. In this case, the shell is experiencing a very weak reaction force for a `long time' before jumping --- when jumping happens otherwise (i.e.~away from the minima of $R(T)$) the force changes rapidly and hence, while $R(T)=0$ denotes the moment of jumping, the shell experiences larger forces closer to jumping.

\begin{figure*}[htbp]
  \centering
     \begin{overpic}[width=0.6\linewidth]{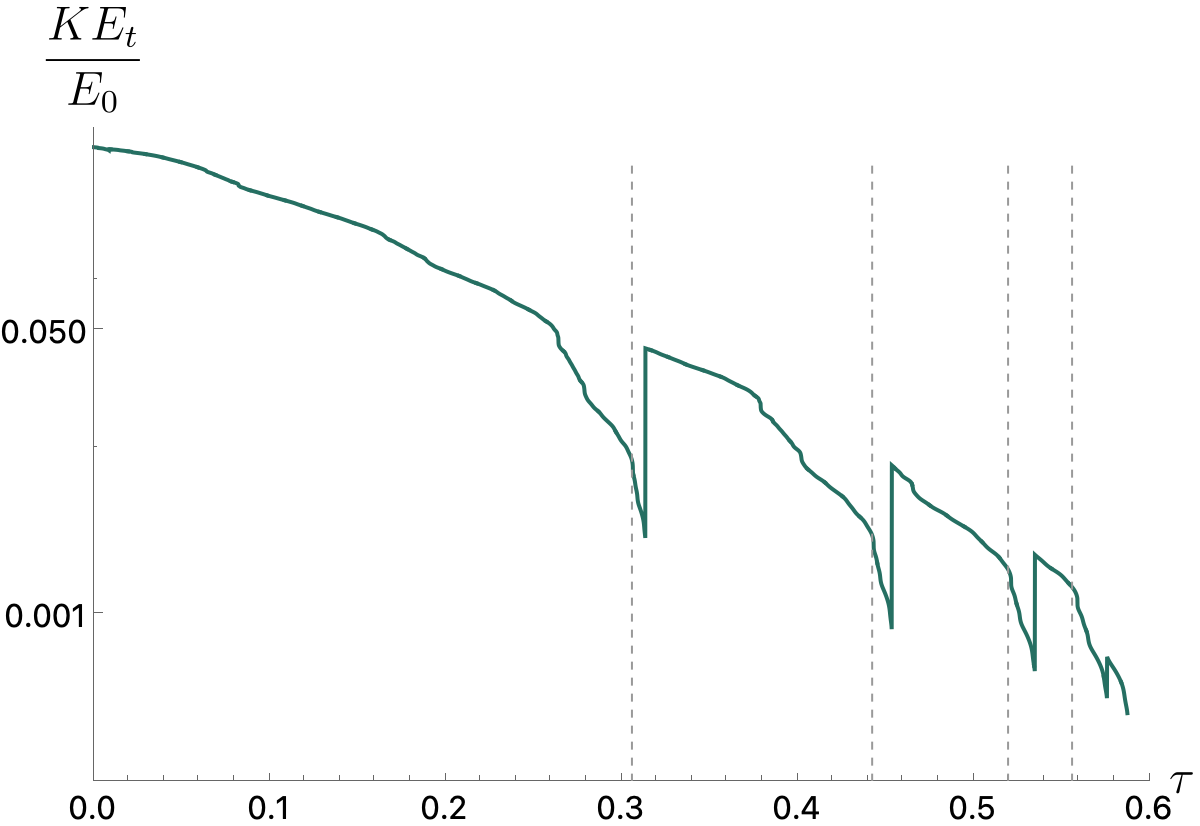}\put(-5,65){\textbf{A}}\end{overpic}
     
       \vspace{4em}
       
     \begin{overpic}[width=0.6\linewidth]{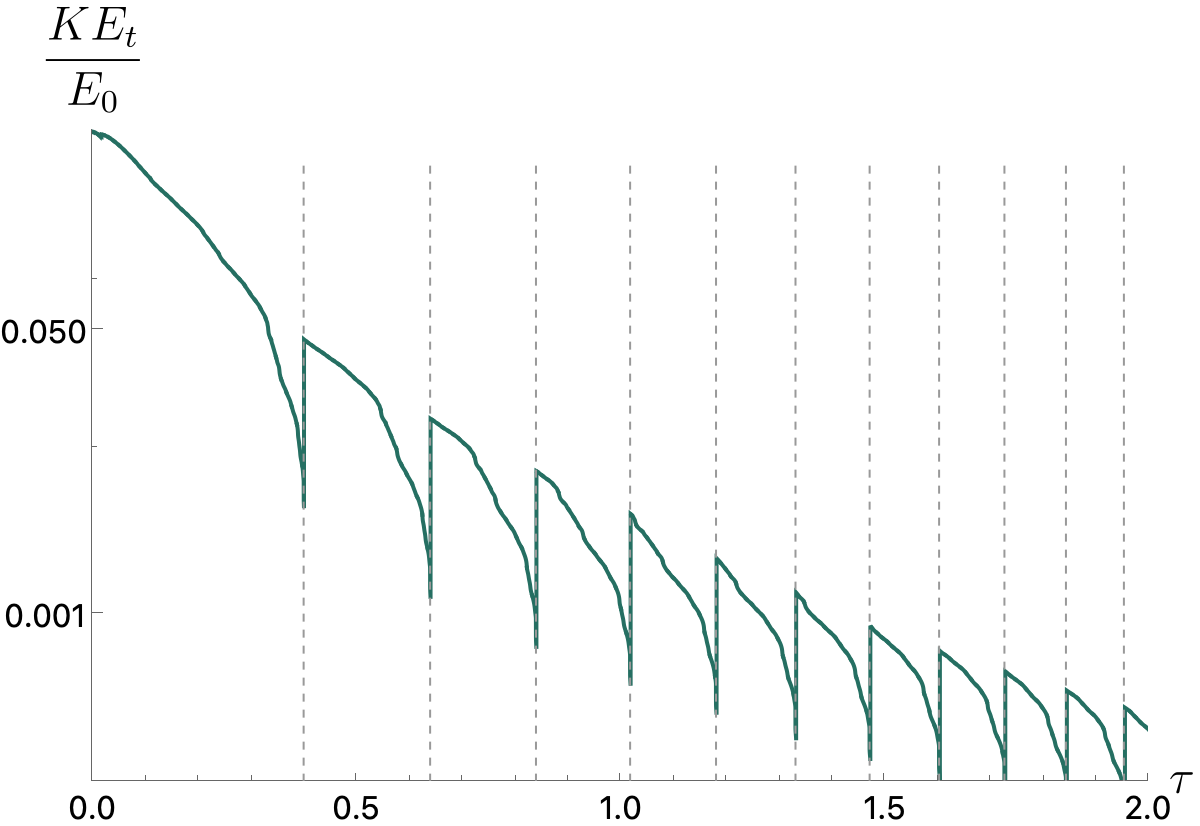}\put(-5,65){\textbf{B}}\end{overpic}
  \caption{Efficiency of jumping for for $\beta=1$ in panel (\textbf{A}), and $\beta=10$ in panel (\textbf{B}); $H=0.1$ for both (\textbf{A}) and (\textbf{B}). Predicted pessimal values of $\tau$ from the numerical roots of \Eq{\ref{eq:pessimal}} are plotted as vertical dashed lines. 
  }
  \label{fig:pessimal}
\end{figure*}

To estimate these pessimal values we need to find the local minima in the oscillating part of $R(T)$ from \Eq{\ref{eq:R}} ($H\tau < 1$). Truncating $R(T)$  at the fundamental mode, the first minimum is located at approximately
\begin{equation}
    T\approx\frac{3}{2}\frac{\pi}{\omega_1}.
\end{equation}
The accuracy of this result increases as $\tau$ increases. The first pessimal value is thus given by the roots of $R(T=\tfrac{3}{2}\tfrac{\pi}{\omega_1})$ --- as given by \Eq{\ref{eq:R}} with the series truncated at the main mode, and considering the asymptotic limit $\tau\gg1$, we find
\begin{align}
    R\left(
    \frac{3}{2}\frac{\pi}{\omega_1}\right)\approx&\frac{1}{2\beta}-\left[\frac{{\rm d}^3 Z_0(X)}{{\rm d} X^3}\right]_{X=1/2^-}^{X=1/2^+}\left\{\frac{-2\pi^4\tau}{\pi(1+\pi^4\tau^2)\omega_1}e^{-\tfrac{3\pi H}{4\omega_1}}\right.\nonumber\\
    &\left.+\frac{1}{2}e^{-\tfrac{3\pi}{2\omega_1\tau}}\frac{\cosh{\frac{(1-H\tau)^{1/4}}{2\sqrt{2\tau}}}\cos{\frac{(1-H\tau)^{1/4}}{2\sqrt{2\tau}}}}{\cosh^2{\frac{(1-H\tau)^{1/4}}{2\sqrt{2\tau}}}\cos^2{\frac{(1-H\tau)^{1/4}}{2\sqrt{2\tau}}}+\sinh^2{\frac{(1-H\tau)^{1/4}}{2\sqrt{2\tau}}}\sin^2{\frac{(1-H\tau)^{1/4}}{2\sqrt{2\tau}}}}\right\}\nonumber\\
    &=0.
    \label{eq:pessimal}
\end{align}
Successive pessimal values are obtained by solving \Eq{\ref{eq:pessimal}} evaluated at $T=\tfrac{7}{2}\tfrac{\pi}{\omega_1},\tfrac{11}{2}\tfrac{\pi}{\omega_1},\tfrac{15}{2}\tfrac{\pi}{\omega_1},\, \dots$. \Fig{\ref{fig:pessimal}} shows these for $\beta=1,10$, and $H=0.1$.

\newpage
\subsection*{\label{sec:larva} Larval Beetles}

\subsubsection*{\label{sec:parameters}Parameter Values}
\begin{figure*}[htbp]
  \centering
      \includegraphics[width=\linewidth]{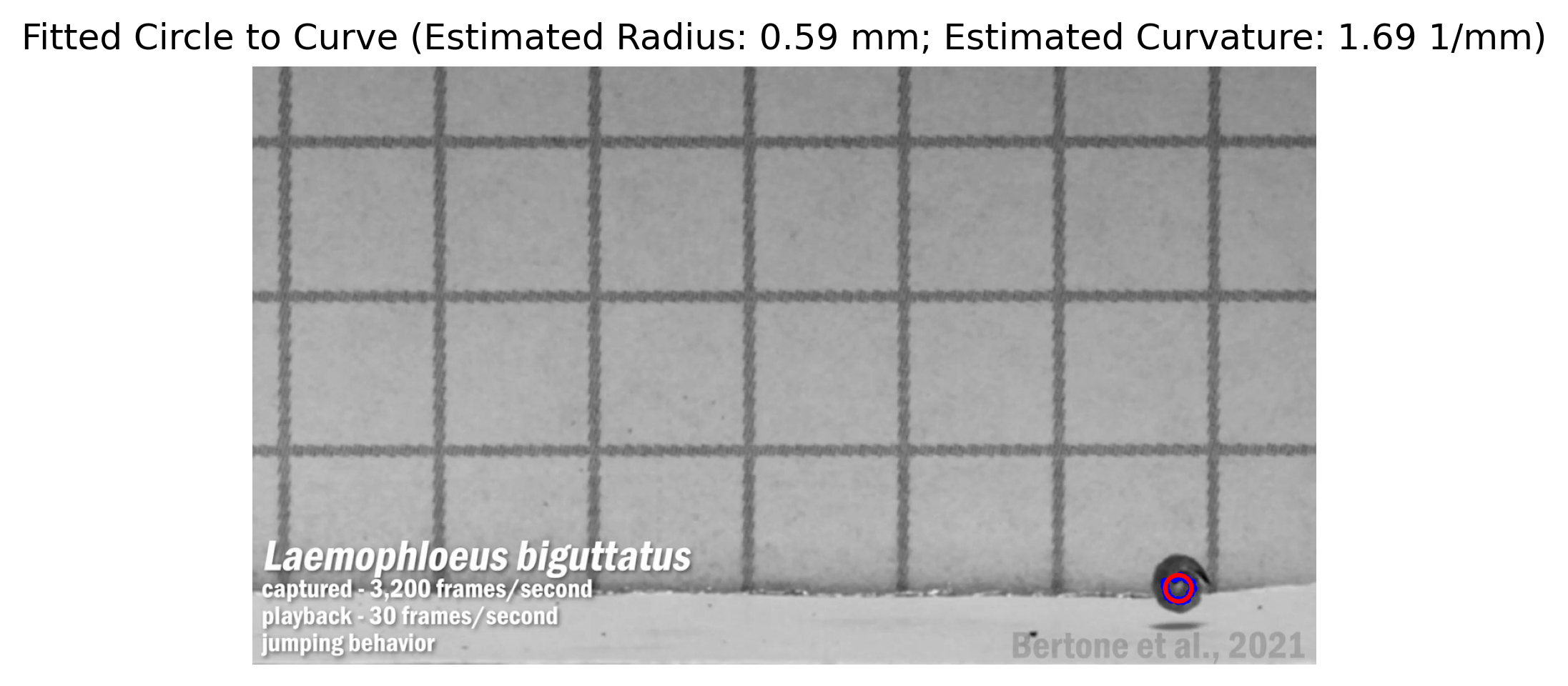}
  \caption{Measurement of the radius of curvature for larval beetles by fitting a circle (in red). Adapted from~\cite{BertoneGibson22}.
  }
  \label{fig:curvature}
\end{figure*}

We measured the radius of curvature of the larval beetles by fitting a circle to their bodies in the video from~\cite{BertoneGibson22}, see \Fig{\ref{fig:curvature}}. We calculate an estimated radius of approximately $0.6~{\rm mm}$ ($|\kappa_0^{\rm max}|\approx 1.69~{\rm mm}^{-1}$).

From these videos, we also measure a curling time scale $t_{\rm curl}\approx 44~{\rm ms}$. We then estimate the bending stiffness $B=EI$ using $t_{\rm curl}$ and the measured curvature~\cite{Callan-JonesBrun12}
\begin{equation}
B=(t_{\rm curl})^{-2}(\kappa_0)^{-4}\rho,
\end{equation}
and obtain an estimate value for the Young's modulus (taking $h\sim 0.6~{\rm mm}$ $\rho\sim mh^{-2}\ell$ with $m\sim 1.3\,10^{-6}~{\rm kg},\,\,\ell\sim 6~{\rm mm}$ from~\cite{BertoneGibson22}) \[E\sim 10~{\rm kPa}.\] Also, we estimate that \[\beta\sim 1\] for these results.

\subsubsection*{\label{sec:energy}Stored Energy}
\begin{figure*}[htbp]
  \centering
    \includegraphics[width=0.6\linewidth]{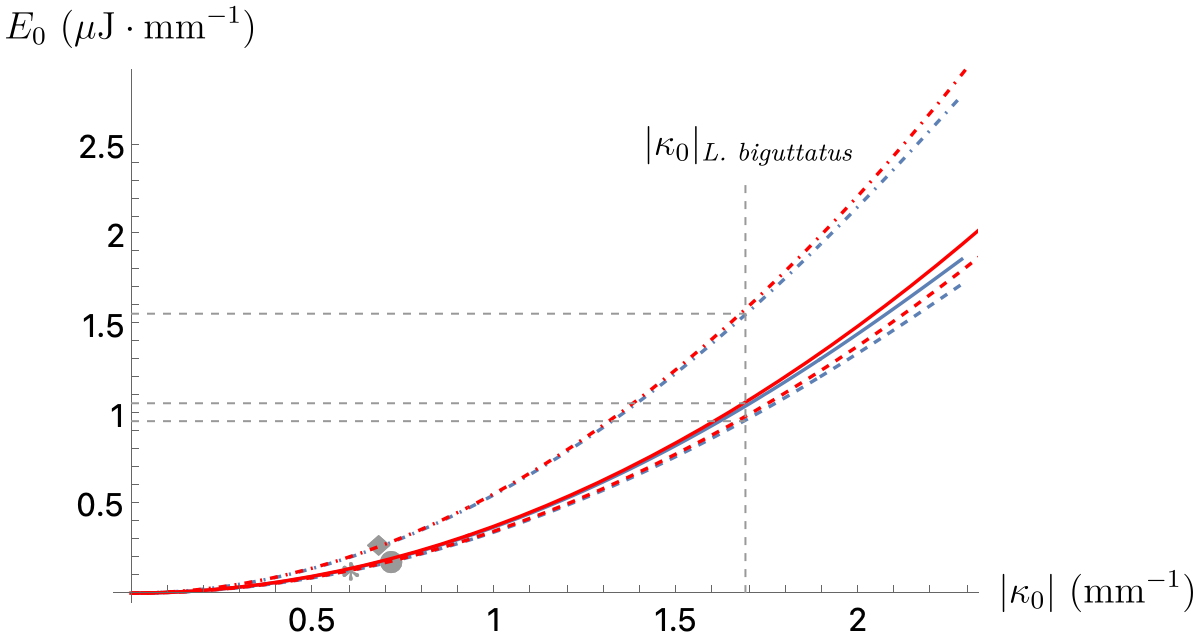}
  \caption{Energy stored (gravitational+bending) of a latched cylindrical shell, with dimensions as per \emph{Laemophloeus biguttatus}~\cite{BertoneGibson22}, as a function of the shell curvature (linearized theory in red and the non-linear model in blue). Gray symbols denote the stored energy per attachment point (or pair of legs for larvae) estimated by the corresponding curvature such that $F_0(\kappa_0)\sim\frac{1}{3}F_0[(\kappa_0)_{\text{\it L. biguttatus}}]$, and assuming the linear theory for $F_0(\kappa_0)$ (star for $E\sim 10~\mathrm{kPa}$, circle for $E\sim 9.25~\mathrm{kPa}$, and diamond for $E\sim 14.9~\mathrm{kPa}$). Horizontal dashed lines show the stored energy at $|\kappa_0|=|\kappa_0|_{\text{\it L. biguttatus}}(=|\kappa_0^{\rm max}|)$ for the different cases.}
  \label{fig:energy}
\end{figure*}

To examine in the main text the attaching force of the larval beetles (see \Fig{3} in the main text, with dimensions as per the results in the previous section). We complement this exploration with the analysis of the stored energy (gravitational+bending) by the larval beetles as a function of the curvature. We plot the results using the linearized model \Eq{\ref{eq:stored_energy}} in \Fig{\ref{fig:energy}}. We also include the results for the non-linear model (see \S\ref{sec:nonlinear}), for which the energy is
\begin{equation}
    E_0(\beta,\ell,\kappa_0)=\frac{EI}{\ell}\left[E_0^{b}(\ell,\kappa_0)+\frac{\kappa_0^2\ell}{\beta}2\int_0^{1/2}{\rm d}S\,Z_0^{L,{\rm nonlinear}}\right];
    \label{eq:stored_energy_nolin}
\end{equation}
with
\begin{align*}
E_0^{b}(\ell,\kappa_0) =\; & \int_0^{0.5} 
\left[ 
\frac{1}{2} \, \Theta(S - S_0^{L}) \left( \frac{{\rm d}\theta_L^-}{{\rm d}S} - \ell\kappa_0 \right)^2 
+ \frac{1}{2} \, \Theta(-S + S_0^{L}) \left( \frac{{\rm d}\theta_L^+}{{\rm d}S} - \ell\kappa_0 \right)^2 
\right] \, dS \\
+ & \int_{0.5}^{1} 
\left[ 
\frac{1}{2} \, \Theta(S - S_0^{R}) \left(  \frac{{\rm d}\theta_R^+}{{\rm d}S} - \ell\kappa_0 \right)^2 
+ \frac{1}{2} \, \Theta(-S + S_0^R) \left(  \frac{{\rm d}\theta_R^-}{{\rm d}S} - \ell\kappa_0 \right)^2 
\right] \, dS
\end{align*} (where $\Theta(x)$ is the Heaviside step function); and
\begin{align*}
Z_0^{L,{\rm nonlinear}}&(S, \theta_0, S_0^{L}, \Lambda_Z^L, \ell, \kappa_0)\equiv \int_0^S{\rm d} S^{\prime} \,\sin{}\theta 
\\
=& \frac{1}{\ell |\kappa_0| \Lambda_Z^L} \Bigg[ 
\sqrt{\Delta_L} \Big(
    {\cal E}\left( \tfrac{\pi - 2\theta_0}{4}, k_L \right) 
    - \Theta(-S + S_0^{L}) \, {\cal E}\left( \tfrac{\pi - 2 \theta_{L}^+(S)}{4}, k_L \right) \\
&\quad + \Theta(S - S_0^{L}) \left( 
    {\cal E}\left( \tfrac{\pi - 2 \theta_{L}^+(S)}{4}, k_L \right) 
    - 2 {\cal E}\left( \tfrac{\pi - 2 \theta_{L}^+(S_0^{L})}{4}, k_L \right) 
\right) \Big) \\
&\quad- \frac{(\kappa_0 \ell)^2 + 2\ell |\kappa_0| \Lambda_Z^L \sin(\theta_0)}{\sqrt{\Delta_L}} \Big( 
    {\cal F}\left( \tfrac{\pi - 2\theta_0}{4}, k_L \right) 
    - \Theta(-S + S_{0}^L) \, {\cal F}\left( \tfrac{\pi - 2 \theta_{L}^+(S)}{4}, k_L \right) \\
&\quad + \Theta(S - S_{0}^L) \left( 
    {\cal F}\left( \tfrac{\pi - 2 \theta_{L}^+(S)}{4}, k_L \right) 
    - 2 {\cal F}\left( \tfrac{\pi - 2 \theta_{L}^+(S_{0}^L)}{4}, k_L \right)
\right) \Big) \Bigg].
\end{align*}
(All notations involved in these expressions are defined in \S\ref{sec:nonlinear}.)

\subsection*{\label{sec:experiments}Experiments with a curved strip}
We observe in our experiments with a curved strip (see \Fig{1C} in the main text) a decay of the oscillations during the airborne phase after jumping.
To estimate the damping $H$ from these oscillations, we approximate the shape of the shell as a parabola, $z(x,t)=a_0\cos{(\omega_0 t)}\,x^2,$
where $a_0$ is the amplitude of the parabola describing the shell at rest (\Fig{\ref{fig:cylindrical_shell}}) and $\omega_0$ is the natural frequency of the shell.
\begin{figure*}[htbp]
  \centering
      \includegraphics[width=0.5\linewidth]{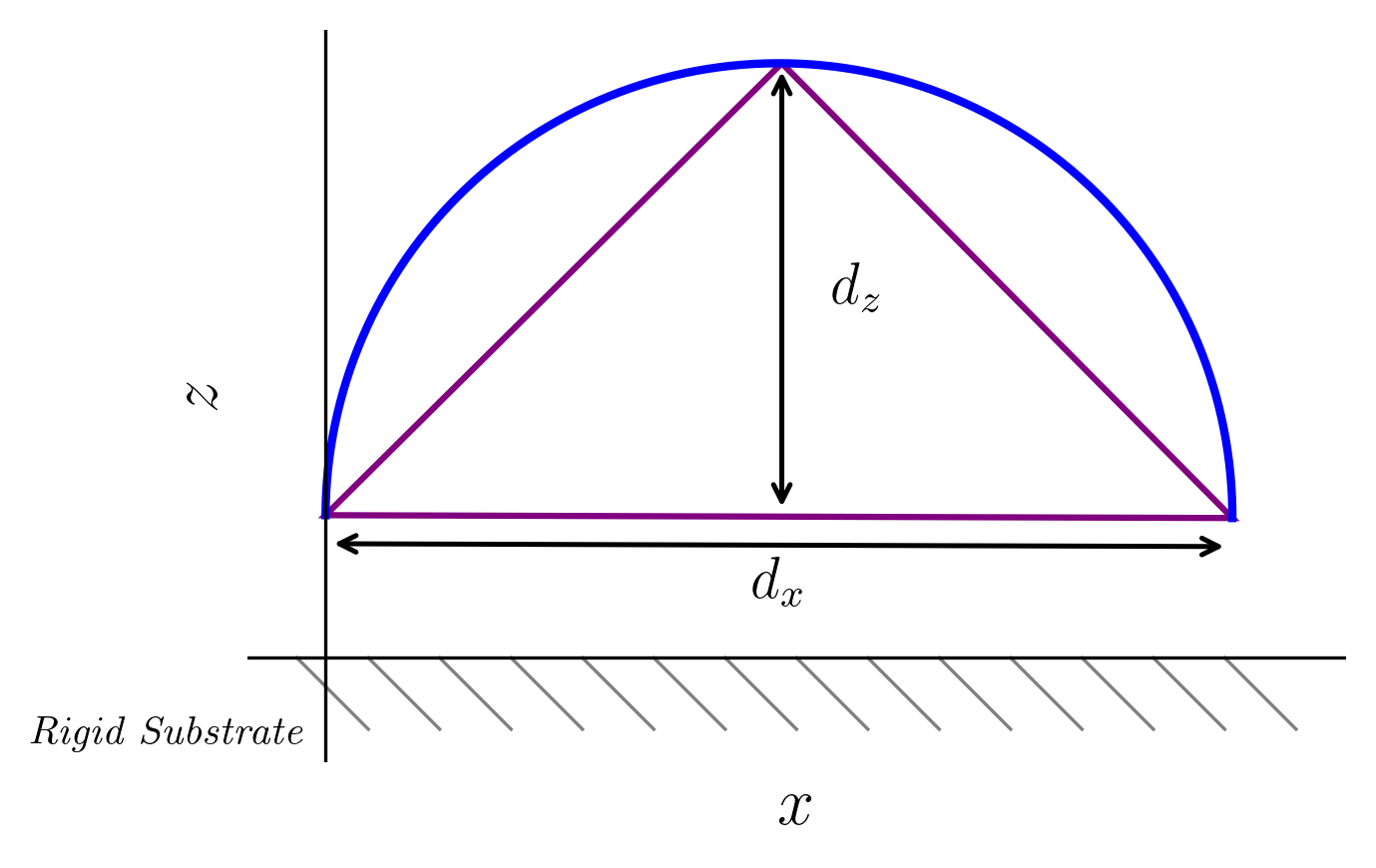}
  \caption{Schematic representation of the cylindrical cap in the airborne phase, with a geometric model of the amplitude.
  }
  \label{fig:triangle}
\end{figure*}
Experimentally, we track three points on the shell: the two ends and the midpoint (see \Fig{\ref{fig:triangle}}). Assuming that the shell is perfectly symmetric, these three points create an isosceles triangle of base $d_x$ and altitude $d_z$. With some trigonometry we then obtain an estimate of the oscillating amplitude of the shell
\begin{equation}
    a(t)=\frac{d_z(t)}{d_x^2(t)}.
\end{equation}

We then fit the oscillations of $a(t)$ with $Z(X,T)$ from the linearized model in \S\ref{sec:model}. $Z(X,T)$ in this phase is described by the dynamic beam equation~\ref{eq:linearized_beam_damped} with free boundaries:
\begin{equation}
    \left.\frac{\partial^2 Z}{\partial X^2}\right|_{X,0,1}=\frac{\kappa_0}{|\kappa_0|}\quad \text{and}\quad \left.\frac{\partial^3 Z}{\partial X^3}\right|_{X,0,1}=0.
\end{equation}

We solve this boundary value problem (neglecting gravity, i.e. $\beta\gg 1$) by letting
\begin{equation}
    Z(X,T)=\zeta(X,T)+w(X),
\end{equation}
where
\begin{equation}
    w(X)=\frac{1}{2}\frac{\kappa_0}{|\kappa_0|}X^2,
\end{equation}
and $\zeta(X,T)$ is the separable solution
\begin{align}
    \zeta(X,T)=&\sum_{n=1}^{\infty} c_1^{(n)}\big[ \big(\cos k_n X+\cosh k_n X\big)-\frac{\cos k_n+\cosh k_n}{\sin k_n-\sinh k_n} \big(\sin k_n X+\sinh k_n X\big)\big]\nonumber\\
    &e^{-\frac{H}{2}T}\left[a_n \cos{(\omega_n T+\varphi_n)}\right];
\end{align}
with \[\omega_n=\frac{H}{2}\sqrt{\frac{4 k_n^4}{H^2}-1},\] and $k_n$ are the roots of the transcendental equation
\begin{equation}
\cos{k}\cosh{k}=1.
\end{equation}

\begin{figure*}[htbp]
  \centering
      \includegraphics[width=0.65\linewidth]{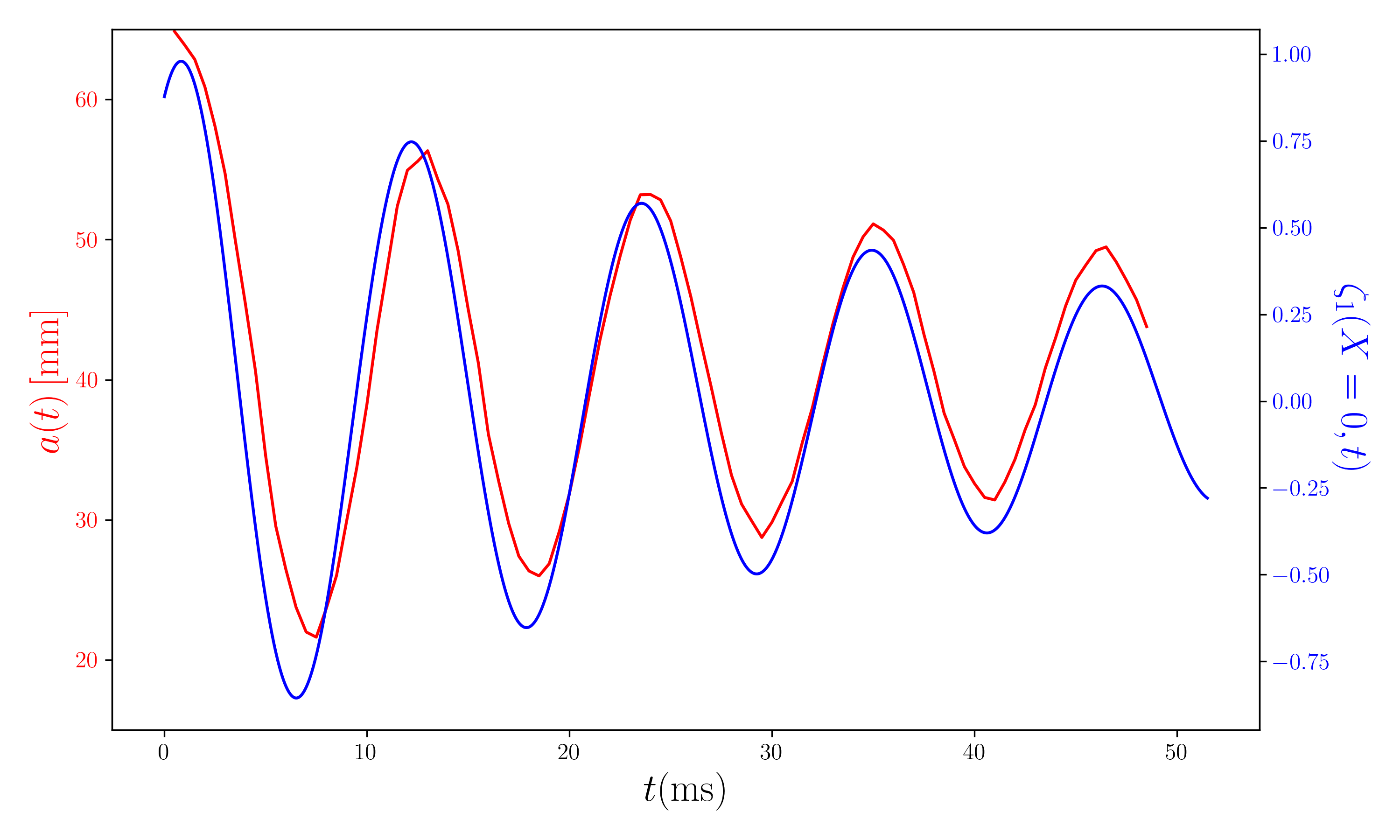}
  \caption{Fitting the amplitude $a(t)$ in the experiments with the fundamental mode of the linearized theory.
  }
  \label{fig:airborne}
\end{figure*}

We fit the decay of the oscillations of $a(t)$ to the fundamental mode \[\zeta_1(X,T)\propto e^{-\frac{H}{2}T}\left[\cos{(\omega_1 T+\varphi_1)}\right].\] Note that we make the time $T$ dimensional with the timescale $(\rho h\ell^4/EI)^{1/2}$, and have \[\rho=1.4\cdot 10^3~{\rm kg}\cdot{\rm m}^{-3},\,h=0.3\cdot 10^{-3}~{\rm m}\,,\ell=6.7\cdot 10^{-2}~{\rm m}\,, \text{~and~} E=2.3\cdot 10^{9}~{\rm Pa}\] in the experiments. The results shown in \Fig{\ref{fig:airborne}} lead to the estimate of $H\approx 2$. For a curvature \[|\kappa_0|\approx \pi/\ell\] as in these experiments, we estimate $\beta\approx 10$.

\section{\label{sec:mass-spring}Simplified 1D  mass-spring system}
\begin{figure*}[htbp]
  \centering
      \includegraphics[width=1\linewidth]{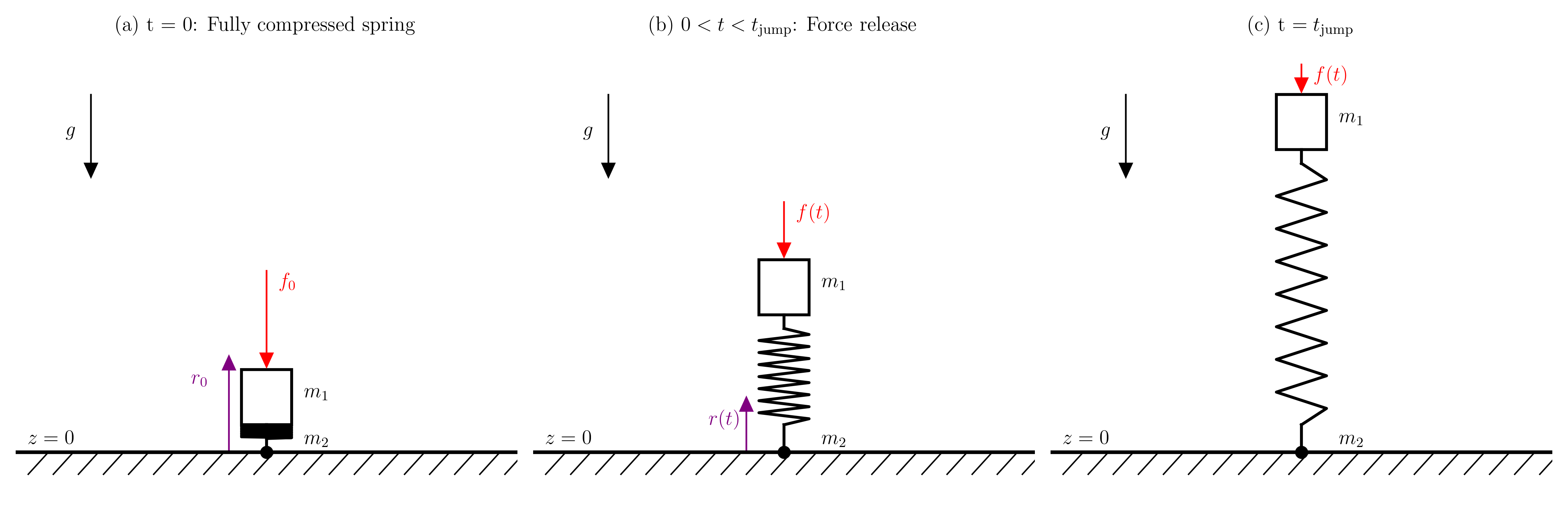}
  \caption{Schematic representation of the mass-spring model of a 1D force-controlled LaMSA mechanism.
  }
  \label{fig:massspring}
\end{figure*}

As a point of comparison for the jumping arch considered in the main paper, we examine here the jumping efficiency of a force-controlled latch mechanism in a 1D  system that consists of two masses ($m_1,m_2$) joined by a spring (with spring constant $k$ and natural length $\ell_0$) --- see \Fig{\ref{fig:massspring}}. This is similar in spirit to the experiments of~\cite{AguilarLesov12}. The system is placed vertically with $m_2$ lying on a flat rigid surface and  subject to gravity. At $t=0$, the spring is fully compressed (i.e.~an external force $f_0= k \ell_0-m_1 g$ is applied onto $m_1$) --- hence the positions of the masses $m_1$ and $m_2$  are equal, $z_1(t=0)=z_2(t=0)=0$.

In the terms of the framework of jumping via a force-controlled LaMSA mechanism, the total force is applied by an external actuator that acts as a latch.
For $t>0$ the initial compression force is relaxed following a certain imposed function of time, $f(t)$. This mimics the force-controlled latch release in the previous curved cylindrical cap (see \S\ref{sec:model}) and the experiments of a elastic strip in the main text.

Prior to jumping, i.e.~for $0<t<t_{\rm{jump}}$, $m_2$ remains on the surface and hence it is subject to a reaction force, $r(t)$, from the substrate; the equations of motion are:
\begin{align}
m_1 \ddot{z_1}(t)&=-\gamma\,\dot{z_1}(t)-f(t)-m_1 g-k\left(z_1(t)-\ell_0\right),~\text{and}\\
0&=-m_2 g+r(t)+k\left(z_1(t)-\ell_0\right),
\label{eq:massspring_balance}
\end{align}
where $\gamma$ is a damping constant.

The force balance for $m_2$ yields the jumping condition --- the smallest $t$ for which $r(t)=0$, which in turn gives
\begin{equation}
z_1(t=t_{\rm{jump}})=\frac{m_2 g}{k}+\ell_0.
\end{equation}
Before solving these equations, we nondimensionalize lengths by $\ell_0$, time by $(m_1/k)^{1/2}$ and forces by $f_0$. Denoting dimensionless versions of the variables using the upper case equivalent, the dimensionless equation of motion is:
\begin{equation}
\ddot{Z_1}(T)+\Gamma\dot{Z_1}(T)+Z_1(T)=(1-1/\beta)(1-F(T));
\label{eq:massspring_Z1}
\end{equation}
where we retained analogous notation as in \S\ref{sec:dissipation} for the dimensionless numbers
\begin{equation}
\beta\equiv\frac{k\ell_0}{m_1 g}\quad \text{and}\quad \Gamma\equiv\frac{\gamma}{(k\, m_1)^{1/2}},
\end{equation}
which measure the relative importance of spring to gravitational forces and the relative importance of damping to spring force, respectively.
Initial and final conditions are
\begin{equation}
Z_1(0)=0,~\dot{Z_1}(0)=0,~Z_1(T=T_{\rm{jump}})=1+\mu/\beta,
\label{eq:masspring_cond}
\end{equation}
where $\mu\equiv m_2/m_1$.

Next, we solve the dimensionless equations of motion with different forms of the controlled force, $F(T)$. Solutions can be analytically computed by a Laplace transform method, for example.

\subsection*{\label{sec:piecewise}Analytical expressions}
The complementary function for \Eq{\ref{eq:massspring_Z1}} shows that  underdamped oscillations are expected for $\Gamma<2$; the frequency of these oscillations is
\begin{equation}
\omega=\sqrt{1-\frac{\Gamma^2}{4}}.
\end{equation}

\begin{figure}[htbp]
  \centering
    \includegraphics[width=0.7\linewidth]{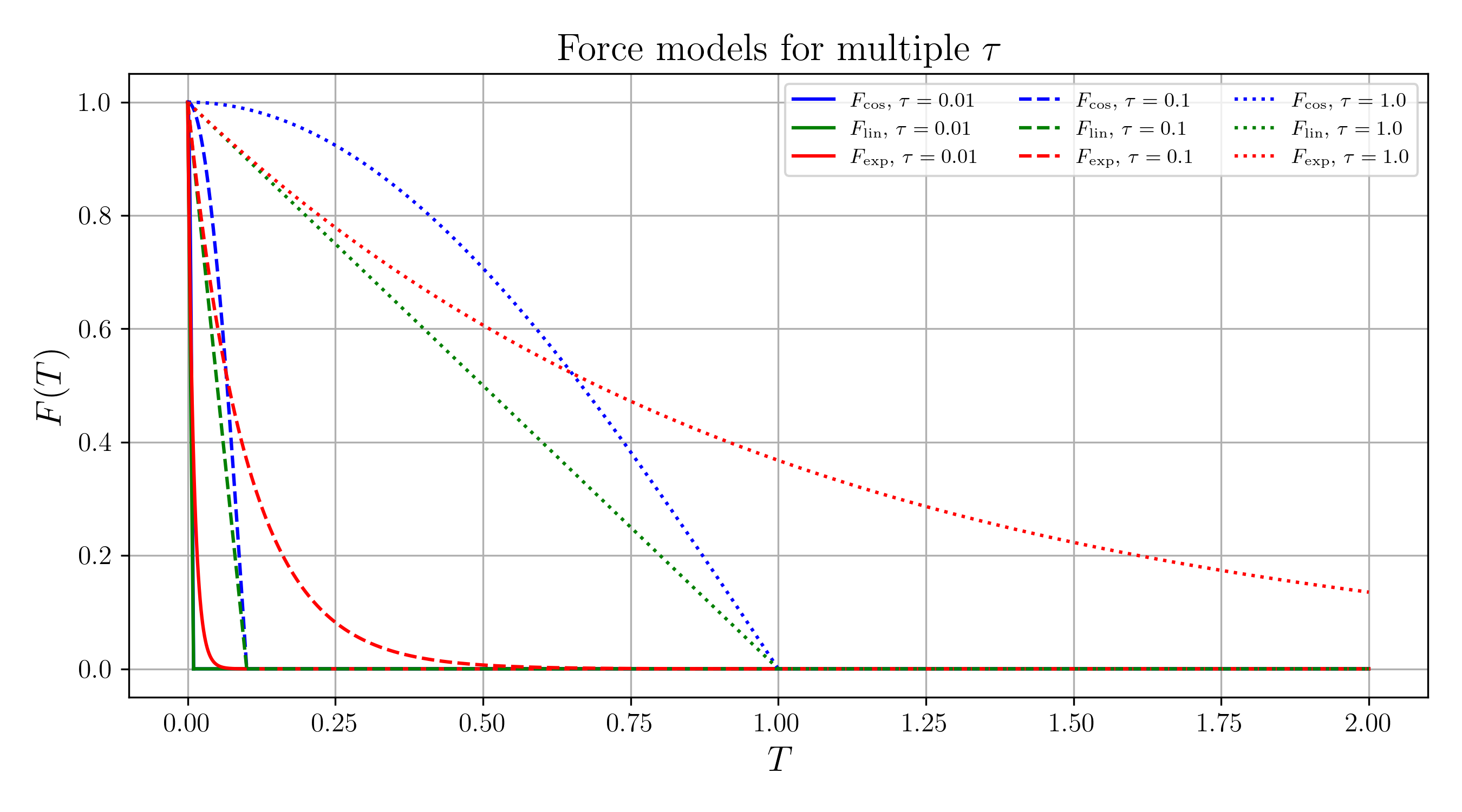}
\caption{Force-controlled latch release following three different loading protocols --- described by \Eq{\ref{eq:relaxation}} --- and with different values of the time constant $\tau$.}
\label{fig:1Dforces} 
\end{figure}

We consider three different functions for the force relaxation, each with relaxation timescale $\tau$:
\begin{align}
F_{\rm{cos}}(T)=\Theta(\tau-T)\cos \left(\frac{\pi}{2\tau}T\right);~ F_{\rm{lin}}(T)=\Theta(\tau-T)\left(1-\frac{T}{\tau}\right);~ F_{\rm{exp}}(T)=\exp(-T/\tau),
\label{eq:relaxation}
\end{align} where $\Theta(\tau-T)$ is a Heaviside function that ensures that for the cosine- and linear-relaxation cases, the force vanishes for times $T\geq\tau$, see \Fig{\ref{fig:1Dforces}}.
In these cases, the outer solution ($T>\tau$, $F(T)=0$) is the homogeneous continuation
\begin{equation}
Z_1(T) =
(1-\beta^{-1})
+ e^{-\frac{\Gamma}{2}(T-\tau)}
\left[
\big(Z_1(\tau)-(1-\beta^{-1})\big)\cos\!\big(\omega (T-\tau)\big)
+ \frac{
\dot{Z_1}(\tau)
+ \frac{\Gamma}{2}\big(Z_1(\tau)-(1-\beta^{-1})\big)
}{\omega}
\sin\!\big(\omega (T-\tau)\big)
\right].
\end{equation}

\subsubsection*{\label{sec:cosine}Cosine relaxation}
Considering a piecewise force with a cosine relaxation for $T\leq\tau$,
\begin{equation}
F_{\cos}(T) =
\begin{cases}
\cos(\omega_0 T), & T \le \tau, \\[6pt]
0, & T > \tau ,
\end{cases}
\label{eq:cosine}
\end{equation}
where $\omega_0 = \frac{\pi}{2\tau}$,  the inner solution ($T\leq \tau$) is
\begin{align}
Z_1^{(\cos)}(T)
&= \Zinf
\Bigg[
1
- e^{-\frac{\Gamma T}{2}}
\left(
\cos\omega T
+ \frac{\Gamma}{2\omega}\sin\omega T
\right)
\nonumber \\[6pt]
&\quad
- \frac{e^{-\frac{\Gamma T}{2}}}{2\omega}
\Big(
S(T)\,\sin\omega T
- C(T)\,\cos\omega T
\Big)
\Bigg],
\qquad T\le\tau ,
\end{align}
 with 
 \begin{align}
S(T)
&=
\sum_{\sigma=\pm}
\frac{
e^{\frac{\Gamma T}{2}}
\left(
2\Gamma \cos b_\sigma T
+ 4 b_\sigma \sin b_\sigma T
\right)
- 2\Gamma
}
{\Gamma^2 + 4 b_\sigma^2},\nonumber
\\[8pt]
C(T)
&=
\sum_{\sigma=\pm}
\frac{
4 b_\sigma
+ 2 e^{\frac{\Gamma T}{2}}
\left(
-2 b_\sigma \cos b_\sigma T
+ \Gamma \sin b_\sigma T
\right)
}
{\Gamma^2 + 4 b_\sigma^2} ;\nonumber\\[8pt]
\text{and}\qquad b_{\pm} &= \omega \pm \omega_0 .\nonumber
\end{align}
The full piecewise solution is
\begin{equation}
Z_1(T)=
\begin{cases}
Z_1^{(\cos)}(T), & T \le \tau, \\[6pt]
\begin{aligned}
\Zinf
&+ e^{-\frac{\Gamma}{2}(T-\tau)}
\Bigg[
\big(Z_1^{(\cos)}(\tau)-\Zinf\big)
\cos\omega (T-\tau)
\\
&\quad
+ \dfrac{
{Z_1^{(\cos)}}'(\tau)
+ \frac{\Gamma}{2}\big(Z_1^{(\cos)}(\tau)-\Zinf\big)
}{\omega}
\sin\omega (T-\tau)
\Bigg],
\end{aligned}
& T > \tau .
\end{cases}
\label{eq:solcos}
\end{equation}

\subsubsection*{\label{sec:linear_rel}Linear relaxation}
Considering a piecewise, linearly-relaxed driving force
\begin{equation}
F_{\mathrm{lin}}(T) =
\begin{cases}
1 - \frac{T}{\tau}, & T \le \tau, \\[2pt]
0, & T > \tau ,
\end{cases}
\label{eq:linear}
\end{equation}
  the inner solution ($T\leq \tau$) is
  \begin{align}
Z_1^{(\mathrm{lin})}(T)
&= \Zinf
\Bigg[
1
- \frac{1 - \frac{T}{\tau} + \frac{\Gamma}{\tau D}}{D}
+ e^{-\frac{\Gamma T}{2}}
\Bigg(
\left(\frac{1 + \frac{\Gamma}{\tau D}}{D}-1\right)\cos\omega T
\nonumber \\[2pt]
&\quad
- \left(
\frac{\Gamma}{2\omega}
+ \frac{-\frac{\Gamma}{2} + \frac{\omega^2 - (\Gamma/2)^2}{\tau D}}{\omega D}
\right) \sin\omega T
\Bigg)
\Bigg],
\qquad T \le \tau;
\end{align} with $D = \omega^2 + \left(\frac{\Gamma}{2}\right)^2$. The full piecewise solution is then
\begin{equation}
Z_1(T) =
\begin{cases}
Z_1^{(\mathrm{lin})}(T), & T \le \tau, \\[4pt]
\begin{aligned}
\Zinf
&+ e^{-\frac{\Gamma}{2}(T-\tau)}
\Bigg[
\big(Z_1^{(\mathrm{lin})}(\tau) - \Zinf\big)\cos\omega (T-\tau)
\\
&\quad
+ \dfrac{
{Z_1^{(\mathrm{lin})}}'(\tau) + \frac{\Gamma}{2}\big(Z_1^{(\mathrm{lin})}(\tau)-\Zinf\big)
}{\omega} \sin\omega (T-\tau)
\Bigg],
\end{aligned}
& T > \tau .
\end{cases}
\label{eq:sollin}
\end{equation}

\subsubsection*{\label{sec:exponential}Exponential relaxation}
For the exponential relaxation --- $F_{\rm{exp}}(T)$ in eq.~\eqref{eq:relaxation} --- the solution is
\begin{equation}
Z_1^{(\exp)}(T)
=
\Zinf
\left[
1
- \frac{e^{-\frac{\Gamma T}{2}}}{\omega^2 + \omega_{0e}^2} 
\Big(
e^{-\omega_{0e} T}
- (1-\omega^2-\omega_{0e}^2)\cos\omega T
+ \frac{\Gamma(\omega^2 + \omega_{0e}^2) + 2\omega_{0e}}{2\omega} \sin\omega T
\Big)
\right],
\label{eq:solexp}
\end{equation}
where $\omega_{0e} = \frac{1}{\tau} - \frac{\Gamma}{2}$.

\subsection*{\label{sec:piecewise_plots}Jumping efficiency}
\begin{figure}[htbp]
  \centering
   \begin{overpic}[width=0.7\linewidth]{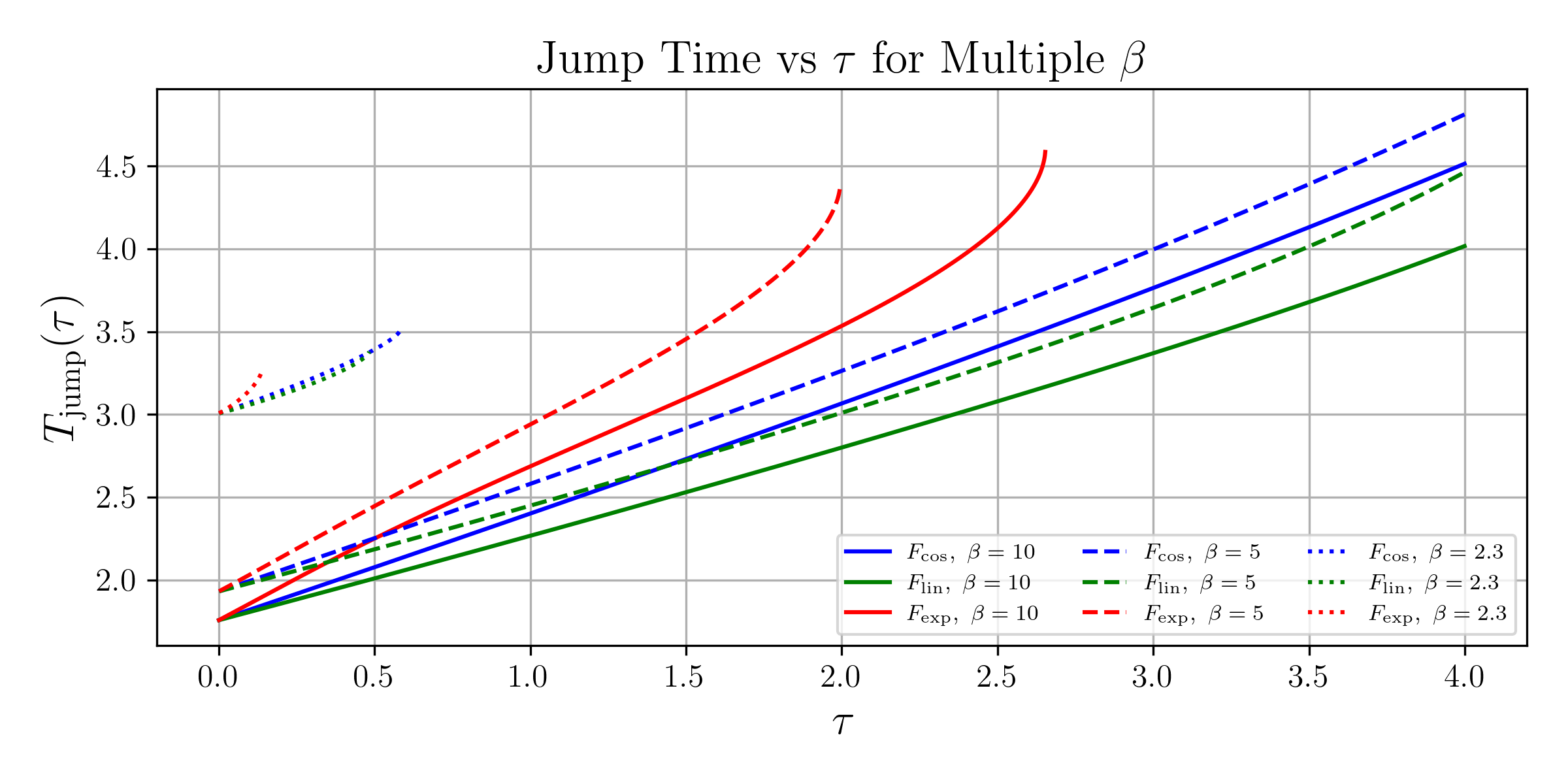}\put(-2,45){\textbf{A}}\end{overpic}
   
     \vspace{4em}

   \begin{overpic}[width=0.7\linewidth]{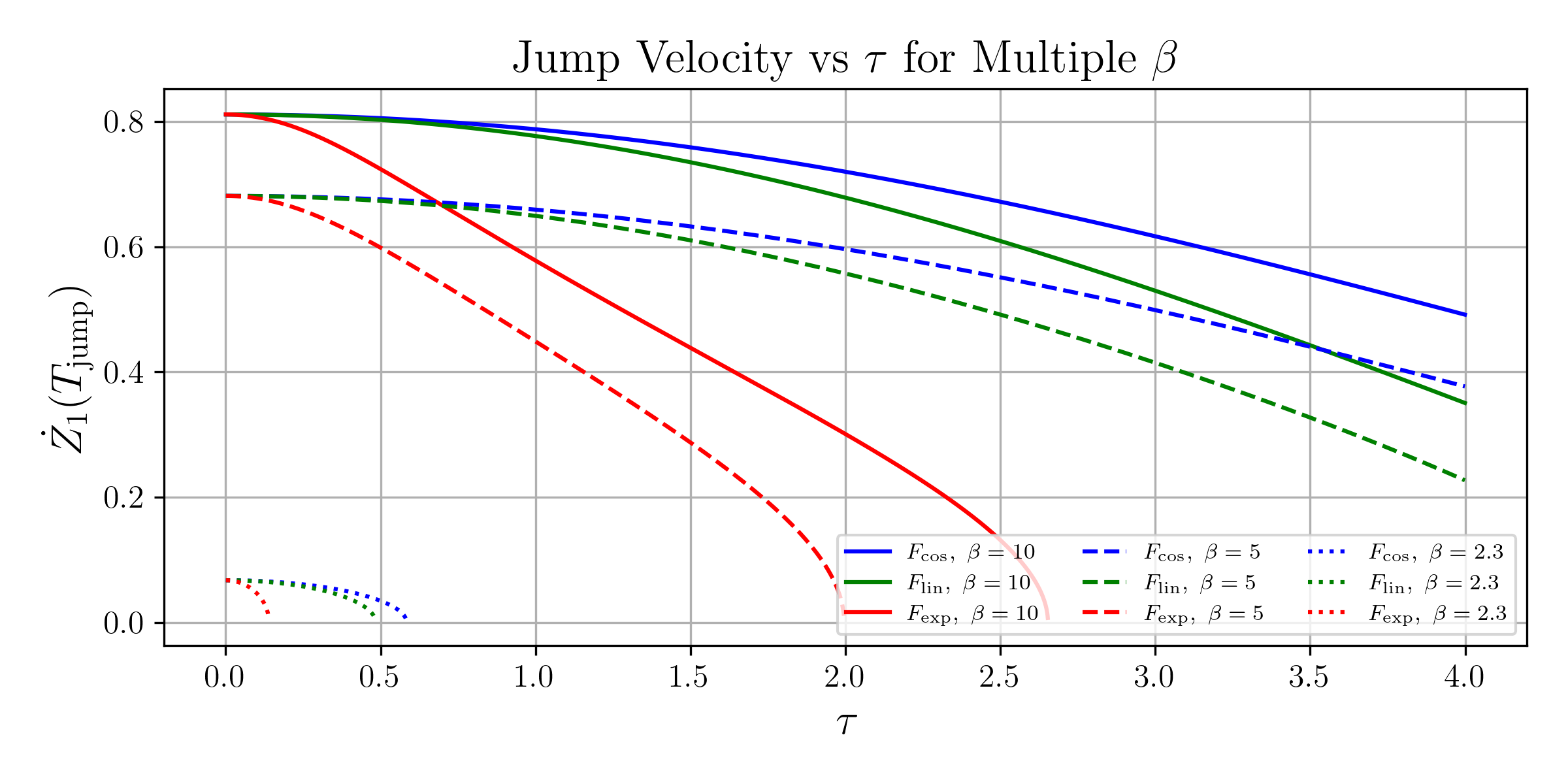}\put(-2,45){\textbf{B}}\end{overpic}
\caption{\label{fig:1Dmassspring} Dependence of the jumping efficiency of a LaMSA mechanism in a simplified 1D mass-spring system for different unloading protocols and time constants $\tau$. (\textbf{A}) Jumping time, defined as the smallest root of $Z_1(T)-(\mu/\beta+1)$ as determined by the \emph{Chebfun} package~\cite{Chebfun} in \emph{Python}. (\textbf{B}) Velocity at the jumping time derived from the analytic solutions \Eqs{\ref{eq:solcos},~\ref{eq:sollin}, \&~\ref{eq:solexp}}.}
\end{figure}

From the analytic solutions \Eqs{\ref{eq:solcos},~\ref{eq:sollin}, \&~\ref{eq:solexp}} we can readily compute the efficiency of jumping in the 1D mass-spring model as a function of the latch-release timescale $\tau$. These results can then be compared to the results  of the continuum model of a curved shell, plotted in \Fig{\ref{fig:pessimal}} and in \Fig{4} in the main text.
We choose $\beta=(2.3,5,10)$, $\Gamma=0.1$ --- comparable to the parameters chosen for \Fig{\ref{fig:pessimal}} and in \Fig{4} of the main text --- and $\mu=0.1$ to compute $T_{\rm{jump}}$ as the smallest root of $Z_1(T)-(\mu/\beta+1)$ (cf.~\Eq{\ref{eq:masspring_cond}}). We plot $T_{\rm{jump}}(\tau)$ in \Fig{\ref{fig:1Dmassspring}A} --- here $T_{\rm jump}$ is determined numerically using the \emph{Chebfun} package~\cite{Chebfun} in \emph{Python}.

Since the jumping efficiency is proportional to the velocity at the jumping time, in \Fig{\ref{fig:1Dmassspring}} we plot $\dot{Z_1}(T=T_{\rm{jump}})$ as a function of the force-control timescale $\tau$.
Notably, the velocity — and hence the efficiency — decays with a long transient linear regime for the exponential force relaxation of a light jumper ($\beta$=5,10).
This behaviour is different from the exponential decay of the efficiency in the continuum model (cf. Fig.~\ref{fig:pessimal} and in \Fig{4} in the main text), which is not observed in any of the 1D simplified mass-spring systems.

\newpage
\bibliography{jumping}